\documentclass[twocolumn,10pt,a4paper]{revtex4-1}
\pdfoutput=1

\usepackage{graphicx}
\usepackage{tikz}
\usepackage{amsmath}
\usepackage{lmodern,microtype}
\usepackage[a4paper,top=20mm,bottom=25mm,left=15mm,right=15mm]{geometry}
\usepackage[utf8]{inputenx}
\usepackage[hidelinks]{hyperref}

\begin{document}

\title{Stochastic model for aerodynamic force dynamics on wind turbine
  blades in unsteady wind inflow} 

\author{M. R. Luhur}
\author{J. Peinke}
\author{M.~K{\"u}hn}
\author{M.~W{\"a}chter}
\affiliation{ForWind, Institute of Physics, Carl von Ossietzky University\\
  26129 Oldenburg, Germany, Email: matthias.waechter@uni-oldenburg.de}	

\begin{abstract} The paper presents a stochastic approach to estimate
  the aerodynamic forces with local dynamics on wind turbine blades in
  unsteady wind inflow. This is done by integrating a stochastic model
  of lift and drag dynamics for an airfoil into the aerodynamic
  simulation software AeroDyn. The model is added as an alternative to
  the static table lookup approach in blade element momentum (BEM)
  wake model used by AeroDyn. The stochastic forces are obtained for a
  rotor blade element using full field turbulence simulated wind data
  input and compared with the classical BEM and dynamic stall models
  for identical conditions. The comparison shows that the stochastic
  model generates additional extended dynamic response in terms of
  local force fluctuations. Further, the comparison of statistics
  between the classical BEM, dynamic stall and stochastic models'
  results in terms of their increment probability density functions
  gives consistent results.

  \vspace{1ex}
  This paper has been published as:\\
  Luhur MR, Peinke J, Kühn M, Wächter M. Stochastic Model for
  Aerodynamic Force Dynamics on Wind Turbine Blades in Unsteady Wind
  Inflow. ASME. J. Comput. Nonlinear Dynam.
  2015;10(4):041010-041010-10. doi:10.1115/1.4028963.
\end{abstract}

\maketitle

\section{Introduction}\label{sec:intro}
Wind energy being safe, significant and fundamental for economical as
well as social development is getting great interest and successfully
penetrating the energy market today. According to Global Wind Energy
Council 2012 projections, the power generation from wind is expected
to reach twice the present global installed capacity
by the end of 2017 \cite{GWEC13}.

In the success of modern wind energy, aerodynamic research has a
significant role \cite{Vermeer03}. The aerodynamic models for wind
turbines are used to study the acting loads on rotor blades in
response to wind inflow. At present, various engineering as well as
computational fluid dynamics (CFD) models exist to foresee the
performance of wind turbines. In computational terms, the
investigations suggest wide choice of engineering methods in
particular the well-known blade element momentum (BEM) method
\cite{Liu12,Ahlund04}. The high fidelity CFD is yet extremely costly
\cite{Liu12} and needs faster and bigger memory computers to achieve
acceptable computational efficiency \cite{Hansen11}. For distributed
loads on wind turbine rotor blades, the aerodynamic models are
combined with dynamic analysis codes such as FAST \cite{Jonkman05},
YawDyn \cite{Laino03}, ADAMS/WT \cite{Laino01}, SIMPACK
\cite{Mulski12}, DHAT \cite{Buhl06}, FLEX5 \cite{Oye99} etc. The
aerodynamic models used by these dynamic analysis codes or other
similar codes are fundamentally based on simple lookup tables, which
contain mean static characteristics for an airfoil at constant angles
of attack (AOAs) \cite{Moriarty05,Weinzierl11} and disregard the
information on system local dynamics. Even the dynamic models
    such as Beddoes-Leishman dynamic stall model use static
    airfoil coefficients which are modified according to AOA and its
    rate of variation and mostly disobey the local force dynamics.

In this contribution, a new concept based on a stochastic approach has
been integrated into the aerodynamic model AeroDyn \cite{Moriarty05}
as an alternative to traditional table lookup method used by the
    classical BEM model. The concept represents a stochastic lift
and drag model, which provides the lift and drag forces with local
dynamics under unsteady wind inflow conditions. The model estimates
the lift and drag coefficients numerically based on the local AOA
\cite{Luhur14}.
The proposed approach thus is a stochastic alternative to the
 classical BEM and Beddoes-Leishman dynamic stall models. 
   
The scope of this paper is to prove the concept by integrating
    the stochastic model into AeroDyn and showing that the newly
    developed concept extracts more load information on rotor blades
    compared to traditional approaches.
A future aim is to achieve a complete stochastic rotor model, which
could provide the full local loading information on blades in a
stochastic sense, leading to an optimum rotor design. Such aerodynamic
model could be combined with a wind energy converter (WEC) model to
obtain a stochastic rotor model. It is necessary to mention here that
this contribution is a primary step towards the final goal and for
simplicity reasons disregards the tower shadow and other related
effects at this stage.

The paper is structured such that Section \ref{sec:AD} provides a
short introduction to the AeroDyn, the input files and the elemental
forces. Section \ref{sec.Mod} introduces the stochastic lift and drag
model. Section \ref{sec.Mod-int-AD} presents the stochastic model
integration into AeroDyn and the results achieved by classical
    BEM, dynamic stall and stochastic model. The final Section
\ref{sec.DisOutlook} summarizes the outcome and outlook of the work.

\section{AeroDyn}\label{sec:AD}
AeroDyn is an add-in software consisting a set of routines to execute
aerodynamic computations for horizontal axis wind turbines (HAWTs). It
can be interfaced with number of dynamic analysis codes in particular
with FAST, SymDyn, YawDyn and ADAMS/WT to carry out the aeroelastic
simulations. These aeroelastic simulation codes differ only in
structural dynamics, the aerodynamic calculations are identical for
them. AeroDyn has no stand alone executable functionality and is
invoked by a dynamic analysis code. However, its separate aerodynamic
forces output file for any element can be obtained by setting an
option from "NOPRINT" to "PRINT" in AeroDyn primary input data file
\cite{Laino02}. It creates one output file in one simulation time for
one selected blade element only.
 
AeroDyn requires information about wind turbine geometry, blade
element velocity, element location, airfoil aerodynamic data,
operating conditions and wind input \cite{Moriarty05}. Based on given
information, it computes the corresponding elemental aerodynamic
forces and delivers to the aeroelastic simulation program to estimate
the distributed forces on wind turbine blades. 
AeroDyn uses different models to perform aerodynamic calculations for
aeroelastic simulations of HAWTs; however, for current computations,
the BEM and dynamic stall models are used. The BEM model is a
well-known classical approach used by different wind turbine designers
with various corrections, whereas the dynamic stall model is based on
the semi-empirical Beddoes-Leishman model which is especially
important for yawed wind turbines. A detailed description of the
classical BEM and dynamic stall models used for present computations
can be found in {\it{Moriarty and Hansen}} \cite{Moriarty05}. The BEM
model is used for estimation of the steady forces (with static table
lookup approach) and the stochastic forces (with stochastic model
addition).

\subsection{AeroDyn input files}\label{subsec-AeroDyn-input}

\begin{figure}
  \center
  \includegraphics[width=0.47\textwidth]{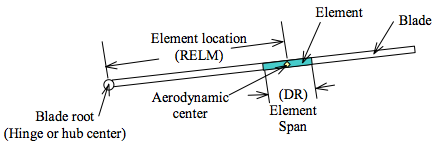}
  \caption{Blade segment nomenclature. Taken from \cite{Laino02}.}
  \label{fig:BEG}
\end{figure}
 
AeroDyn requires three input files to perform the aerodynamic
calculations. These are, the primary data file, the airfoil data file
and the wind file. Latter two are called through paths provided in
primary data file. The primary data file consists of different options
of models for calculation of flow influence. Additionally, it carries
the information on wind reference height (hub height), convergence
tolerance for induction factors, air density, kinematic air viscosity,
time interval for aerodynamic calculations, and the number of blade
elements per blade. For each blade element it is provided with the
location, twist, span and chord length. For more clarity of the
element related parameters; see blade segment terminology given in
Figure \ref{fig:BEG}.

The airfoil data file contains two dimensional static airfoil
characteristics. It carries lift, drag and pitching moment (optional)
coefficients for range of AOAs. Besides, it comprises some additional
parameters pertaining to dynamic stall model. The airfoil
characteristics for intermediate AOAs are obtained by linear
interpolation.

The wind files in AeroDyn are used in two formats, one the hub-height
wind and other the full field turbulence, which are created by either
measurements or simulations. The hub-height wind files are the simple
ones containing either steady or time varying wind data. The full
field turbulence is generated by TurbSim program
\cite{Kelley07,Jonkman12}, which creates two files, one the binary
wind data file and other the summary file. The created wind data
corresponds to all three wind components changing in time and space.
The wind is sampled at frequency of 20 Hz and the turbulence is
generated by a square grid spread over the whole rotor area. The
velocity components at each point of the square grid are provided as
function of time. The components of velocity at each blade element are
obtained with linear interpolation (in terms of averaging or
smoothing) in time and space \cite{Moriarty05}. For more details; see
\cite{Kelley07,Jonkman12}.

\subsection{Elemental forces}\label{subsec.ElementForces}

\begin{figure}
  \center
  \includegraphics[width=0.49\textwidth]{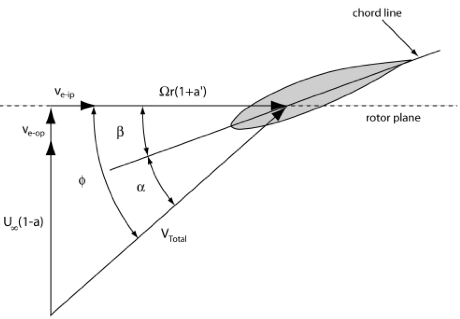}
  \put(-48,50){(a)}
  
  \includegraphics[width=0.29\textwidth]{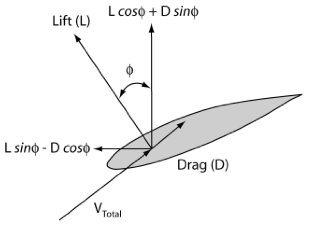}
  \put(0,40){(b)}
  \caption{Scheme of force components on blade section. Angles are
    related to the plane of rotation. (a) Local velocities and flow
    angles on blade element and (b) local forces on blade element.
    Taken from \cite{Moriarty05}.}
  \label{fig:Belm}
\end{figure}

To determine the elemental forces on a rotating blade, the BEM method
is applied on blade section as shown in Figure \ref{fig:Belm}. The
illustration shows the local velocities with flow angles and the
acting forces on the element.

In Figure \ref{fig:Belm}(a), $\phi$ represents the flow angle,
$V_{\mathit{Total}}$ the relative speed, $\alpha$ the AOA and $\beta$
the combination of pitch and twist angles. The flow angle $\phi$ is
the angle between the relative speed and the plane of rotation,
whereas the AOA $\alpha$ is the angle between the relative speed and
the chord of the blade element. The parameter $U_{\infty}$ denotes the
free stream wind velocity, $\Omega$ the blade rotational speed and $r$
the local radius of the blade element. The variables $v_\mathit{e-op}$
and $v_\mathit{e-ip}$ are the out-of-plane and in-plane element
velocities, respectively, originating from blade structural
deflections under pronounced rotation. When the blade rotational speed
is very small, the latter velocities are ignored.

The terms $U_{\infty}(1-a)$ and $\Omega r(1+\acute{a})$ are the
effective axial wind and tangential blade speeds, respectively. The
parameters $a$ and $\acute{a}$ are the axial and tangential induction
factors, where $a$ represents the amount of reduction in axial wind
speed when approaching the blade and $\acute{a}$ the amount of
rotational acceleration to the blade caused by induced wake rotation
opposite to the rotor rotation \cite{Mendez06}.

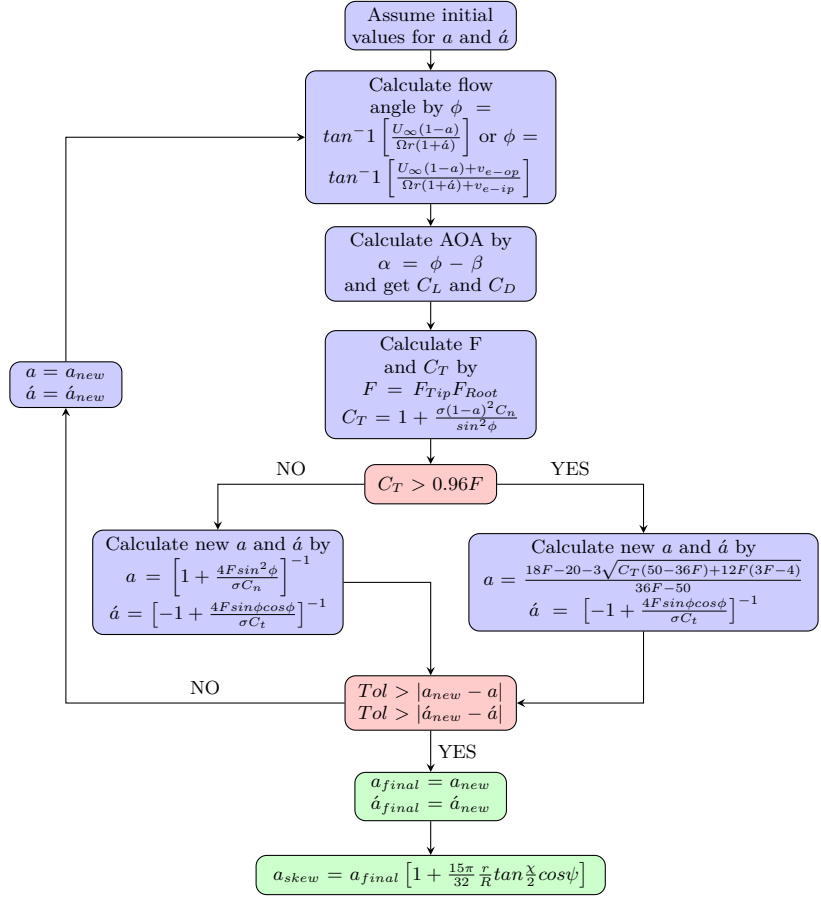
\begin{figure*}[t]
  \center
  
  \begin{tikzpicture}[scale=0.8, node distance = 1.5cm, auto,transform shape]
    
    \tikzstyle{block} = [rectangle, draw, fill=blue!20,text width=8em, text centered, rounded corners, minimum height=2em]
    \node [block] (init) {Assume initial values for $a$ and $\acute{a}$};
    
    \tikzstyle{block} = [rectangle, draw, fill=blue!20, text width=12em, text centered, rounded corners, minimum height=2em]
    \node [block, below of=init,node distance = 1.85cm] (identify) {Calculate flow angle by $\phi=tan^-1\left[\frac{U_{\infty}(1-a)}{\Omega r(1+\acute{a})}\right]$ or 
      $\phi=tan^-1\left[\frac{U_{\infty}(1-a)+v_{e-op}}{\Omega r(1+\acute{a})+v_{e-ip}}\right]$};
    
    \tikzstyle{block} = [rectangle, draw, fill=blue!20, text width=10em, text centered, rounded corners, minimum height=2em]
    \node [block, below of=identify,node distance = 2.1cm] (evaluate1) {Calculate AOA by\\ $\alpha=\phi-\beta$ \\and get $C_L$ and $C_D$};
    
    \tikzstyle{block} = [rectangle, draw, fill=blue!20, text width=10em, text centered, rounded corners, minimum height=2em]
    \node [block, below of=evaluate1,node distance = 2cm] (evaluate2) {Calculate F and $C_T$ by \\ $F=F_{Tip}F_{Root}$ \\ $C_T=1+\frac{\sigma(1-a)^2C_n}{sin^2\phi}$};
    
    \tikzstyle{block} = [rectangle, draw, fill=red!20, text width=6em, text centered, rounded corners, minimum height=2em]
    \node [block, below of=evaluate2,node distance = 1.65cm] (decide) {$C_T>0.96F$};
    
    \tikzstyle{block} = [rectangle, draw, fill=blue!20, text width=12em, text centered, rounded corners, minimum height=2em]
    \node [block, left of=decide,node distance=3.5cm,yshift=-5em] (a-BEM) {Calculate new $a$ and $\acute{a}$ by\\ $a=\left[1+\frac{4Fsin^2\phi}{\sigma C_n}\right]^{-1}$\\$\acute		{a}=\left	[-1+\frac{4Fsin\phi cos\phi}	{\sigma C_t}\right]^{-1}$};
    
    \tikzstyle{block} = [rectangle, draw, fill=blue!20, text width=17em, text centered, rounded corners, minimum height=2em]
    \node [block, right of=decide, node distance=3.5cm,yshift=-5em] (a-Glauert) {Calculate new $a$ and $\acute{a}$ by\\ $a=\frac{18F-20-3\sqrt{C_T(50-36F)+12F(3F-4)}}				{36F-50}$	\\$\acute{a}=\left[-1+\frac{4Fsin\phi cos\phi}{\sigma C_t}\right]^{-1}$};
    
    \tikzstyle{block} = [rectangle, draw, fill=red!20, text width=8em, text centered, rounded corners, minimum height=2em]
    \node [block, below of=decide, node distance=2.65cm,yshift=-3em] (Tol) {$Tol>|a_{new}-a|$ $Tol>|\acute{a}_{new}-\acute{a}|$};
    
    \tikzstyle{block} = [rectangle, draw, fill=blue!20, text width=5em, text centered, rounded corners, minimum height=2em]
    \node [block, left of=evaluate2, node distance=6cm] (update) {$a=a_{new}$\\$\acute{a}=\acute{a}_{new}$};
    
    \tikzstyle{block} = [rectangle, draw, fill=green!20, text width=7em, text centered, rounded corners, minimum height=2em]
    \node [block, below of=Tol, node distance=1.55cm] (stop) {$a_{final}=a_{new}$\\$\acute{a}_{final}=\acute{a}_{new}$};
    
    \tikzstyle{block} = [rectangle, draw, fill=green!20, text width=17em, text centered, rounded corners, minimum height=2em]
    \node [block, below of=stop, node distance=1.3cm] (end) {$a_{skew}=a_{final}\left[1+\frac{15\pi}{32}\frac{r}{R}tan\frac{\chi}{2}cos\psi\right]$};
    
    \tikzstyle{line} = [draw, -stealth]
    \path [line] (init) -- (identify);
    \path [line] (identify) -- (evaluate1);
    \path [line] (evaluate1) -- (evaluate2);
    \path [line] (evaluate2) -- (decide);
    
    \path [line] (decide) -| node[yshift=1.5em,near start] {NO} (a-BEM);
    \path [line] (decide) -| node[yshift=0.1em,near start]{YES} (a-Glauert);
    
    \path [line] (a-BEM) -| (Tol);
    \path [line] (a-Glauert) |- (Tol);
    
    \path [line] (Tol) -| node[yshift=1.5em,near start] {NO} (update);
    \path [line] (update) |- (identify);
    
    \path [line] (Tol) -- node {YES}(stop);
    \path [line] (stop) -- (end);
    
  \end{tikzpicture}
  \caption{Flow chart to iterate for induction factors.}
  \label{fig:Flowchart-BEM}
\end{figure*}

The induction factors are estimated using an iterative process
described in the flow chart Figure~\ref{fig:Flowchart-BEM}. After
initialization, the algorithm iteratively finds values fulfilling the
condition expressed for $\mathit{Tol}$. The parameter $\mathit{Tol}$
is an acceptable tolerance allowed around the true values of axial and
tangential induction factors. The process repeats for each element.
Once the induction factors, inflow angles and AOAs converged to their
final values, the acting forces are estimated for the element.

In flow chart Figure~\ref{fig:Flowchart-BEM}, the function $F$
represents the tip and root loss correction factors in
combination, $C_T$ the thrust loading on the element, $\sigma$ the
local solidity of the rotor, $C_n$ the normal force coefficient
and $C_t$ the tangential force coefficient. The parameter
$a_{\mathit{skew}}$ is the local element induction factor for
skewed wake, $R$ the blade radius, $\chi$ the wake angle and
$\psi$ the blade azimuth angle\footnote{The azimuth angle
  describes the blade angular position in one cycle measured in
  clockwise direction such that it is zero when the blade is
  pointing vertically downwards.}. For more details and derivations of
the relations given in flow chart Figure~\ref{fig:Flowchart-BEM}; see
\cite{Manwell09,Mendez06,Glauert35,Buhl05,Glauert26A,Moriarty05,Glauert26B,Pitt81,Coleman45,Burton11}.

\section{Stochastic lift and drag model}\label{sec.Mod}

A stochastic model of the lift and drag dynamics has been developed
for an advanced characterization of forces on wind turbine airfoils
under unsteady wind inflow. The model parameters are derived from
dynamic measurements of lift and drag forces performed for FX
79-W-151A airfoil in a wind tunnel. The turbulent inflow with
intensity of 4.6\% was generated using a fractal square grid, which
produces the flow with typical intermittent velocity fluctuations
commonly known for free field wind situations; see \cite{Luhur14}. The
model extracts most of the information available in the system
dynamics in terms of a dynamic response. The model reads
\cite{Luhur14}

\begin{multline}
  X_{model}(k)= \\
  X_{langevin}(k)+A\sin\left(\frac{2\pi k}{T}\right)\exp\left[\left(\frac{-k'}{k_o}\right)^S\right],
  \label{eq:Ext mod}
\end{multline}

where the first part of the equation represents the basic model and
the second the extension which accounts for the oscillation effects
with amplitude modulation (breathing) contained in the lift and drag
time series. The parameter $X$ represents the lift and drag
coefficients, $A$ the constant to fix the oscillation amplitude for
lift and drag coefficients, $k$ the discrete time variable and $T$ the
most dominant oscillation period of lift and drag coefficients. The
exponential function in equation (\ref{eq:Ext mod}) controls the
breathing of oscillation along the lift and drag time series, where
$k_o$ is half the average breathing length, $k'=(k~\mbox{mod}~k_o)$
and $S$ is described as
\begin{equation}
  \begin{aligned}
    S = \left\{
      \begin{array}{lr}
        +1,~\mbox{for } (2n)k_o< k \leq(2n+1)k_o\\ \\
        -1, ~\mbox{for } (2n+1)k_o< k\leq 2(n+1)k_o\;,
        \label{eq:S}
      \end{array}
    \right.
  \end{aligned}
\end{equation}	
where $n=0,1,2, \ldots$.

The basic model $X_{langevin}(k)$ in equation (\ref{eq:Ext mod})
corresponds to a first order stochastic differential equation termed
as Langevin equation, cf.\cite{Risken96}, expressed in discrete form as
\begin{multline}
  X(k+1)= \\
  X(k)+\tau D^{(1)}(X,\alpha)+\sqrt{\tau\tilde{F}D^{(2)}(X,\alpha)}~\Gamma(k),
  \label{eq:Langevin}
\end{multline}
where $D^{(1)}(X,\alpha)$ and $D^{(2)}(X,\alpha)$ are the drift and
diffusion functions, also known as first and second Kramers-Moyal
coefficients. Here $\tau$ is the integration time step and $\alpha$ the mean AOA that varies slowly
compared to the fluctuations caused by turbulent inflow conditions. The
parameter $\tilde{F}$ is the correction factor for diffusion function
to incorporate the model extension and $\Gamma(k)$ the Gaussian white
noise termed as Langevin force \cite{Risken96} with mean value
$\langle\Gamma(k)\rangle=0$ and variance
$\langle\Gamma^2(k)\rangle=2$.

The $D^{(1)}(X,\alpha)$ reflects the deterministic part of the system,
whereas $D^{(2)}(X,\alpha)$ quantifies the amplitude of the stochastic
fluctuations. These functions for lift and drag coefficients are
parameterized as \cite{Luhur14}
\begin{equation}
  D^{(1)}(X,\alpha)=m(X-X_o),
  \label{eq:D1-PMTZ}
\end{equation}
\begin{equation}
  D^{(2)}(X,\alpha)=\beta.
  \label{eq:D2-PMTZ}
\end{equation}
Here $m$ is the slope of drift function, $X$ the lift or drag
coefficient, $X_o$ the stable fix point in $X$ where drift function is
zero and $\beta$ the constant diffusion function. The optimized values
for these parameters as function of mean AOAs are given in Tables
\ref{Table:CL-parametrs} and \ref{Table:CD-parametrs}. The
intermediate values can be obtained by linear interpolation. For
    more details and validation of the model; see \cite{Luhur14}.
\begin{table*}[t] 
  \caption{$C_L$ model parameters for AOAs $0^\circ$ to $30^\circ$. The parameter $m$ is the slope of drift function,  $C_{Lo}$ the stable fix point, $\beta$ the optimized diffusion 	function, $\tilde{F}$ the correction factor for diffusion function to incorporate the model extension, $A$ the constant to fix the oscillation amplitude, $T$ the most dominant oscillation period 	and  	$k_o$ half the average breathing length.}
  \center
  \scriptsize
  \begin{tabular}{| c | c | c | c | c | c | c | c || c | c | c | c | c | c | c | c |}
    \hline
    AOA 		& 	$m$   	& 	$C_{Lo}$ 	&	$\beta$		&  	$F$		& 	$A$     		&   	$T$		&   	$k_o$	&	AOA 		& 	$m$   	& 	$C_{Lo}$ 	&	$\beta$		&  	$F$		& 	$A$     		&   	$T$		&   	$k_o$\\\hline
    $0^\circ$ 	& 	$-0.160$ 	& 	$0.228$	&  	$2.50e-04$ 	&  	$0.329$ 	&  	$0.092$ 		& 	$30$		&   	$300$    	&	$16^\circ$ & 	$-0.115$ 	& 	$1.286$	&  	$6.58e-04$ 	&       $0.500$   &      $0.154$         	&       $30.25$    &      $300$\\\hline
    $1^\circ$ 	& 	$-0.160$ 	& 	$0.355$	&  	$3.17e-04$ 	&  	$0.329$ 	&  	$0.105$ 		& 	$30$		&   	$300$    	&	$17^\circ$ & 	$-0.118$ 	& 	$1.278$	&  	$9.21e-04$ 	&  	$0.500$ 	&  	$0.175$ 		& 	$30.5$	&   	$305$\\\hline
    $2^\circ$ 	& 	$-0.157$ 	& 	$0.450$	&       $3.32e-04$      &       $0.330$   	&      $0.109$           	&       $30$         &       $300$       &	$18^\circ$ & 	$-0.121$ 	& 	$1.270$	&  	$1.26e-03$ 	&  	$0.500$ 	&  	$0.199$ 		& 	$30.75$	&   	$300$\\\hline
    $3^\circ$ 	& 	$-0.156$ 	& 	$0.547$	&       $3.23e-04$      &       $0.329$   	&      $0.107$        	&       $30$         &       $300$      &	$19^\circ$ & 	$-0.120$ 	& 	$1.249$	&  	$1.81e-03$ 	&       $0.330$   &      $0.290$            	&       $31.25$   &       $300$ \\\hline
    $4^\circ$ 	& 	$-0.150$ 	& 	$0.641$	&       $2.90e-04$      &       $0.329$   	&      $0.101$              &       $30$        &        $300$       &	$20^\circ$ & 	$-0.125$ 	& 	$1.238$	&  	$2.25e-03$ 	&  	$0.330$ 	&  	$0.318$ 		& 	$31.33$	&   	$300$\\\hline
    $5^\circ$ 	& 	$-0.155$ 	& 	$0.735$	&  	$3.31e-04$ 	&  	$0.329$ 	&  	$0.107$ 		& 	$30$		&   	$300$    	&	$21^\circ$ & 	$-0.130$ 	& 	$1.218$	&  	$3.03e-03$ 	&  	$0.330$ 	&  	$0.348$ 		& 	$32.5$	&   	$300$ \\\hline
    $6^\circ$ 	& 	$-0.154$ 	& 	$0.827$	&  	$3.59e-04$ 	&       $0.330$   &      $0.113$            	&       $30$         &       $300$       &	$22^\circ$ & 	$-0.129$ 	& 	$1.199$	&  	$3.51e-03$ 	&  	$0.330$ 	&  	$0.378$ 		& 	$32.5$	&   	$300$ \\\hline
    $7^\circ$ 	& 	$-0.147$ 	& 	$0.912$	&  	$3.51e-04$ 	&  	$0.330$ 	&  	$0.113$ 		& 	$30$		&   	$300$    	&	$23^\circ$ & 	$-0.133$ 	& 	$1.184$	&  	$4.41e-03$ 	&  	$0.330$ 	&  	$0.415$ 		& 	$32.5$	&   	$300$\\\hline	
    $8^\circ$ 	& 	$-0.142$ 	& 	$0.996$	&  	$3.50e-04$ 	&  	$0.330$ 	&  	$0.115$ 		& 	$30$		&   	$300$    	&	$24^\circ$ & 	$-0.136$ 	& 	$1.164$	&  	$5.26e-03$ 	&       $0.329$   &      $0.450$            	&       $32.33$   &       $300$\\\hline
    $9^\circ$ 	& 	$-0.143$ 	& 	$1.069$	&  	$3.37e-04$ 	&       $0.330$   &      $0.114$            	&       $30$         &       $300$      &	$25^\circ$ & 	$-0.136$ 	& 	$1.146$	&  	$5.39e-03$ 	&  	$0.330$ 	&  	$0.460$ 		& 	$32.35$	&   	$300$\\\hline
    $10^\circ$ & 	$-0.140$ 	& 	$1.141$	&  	$3.05e-04$ 	&  	$0.329$ 	&  	$0.108$ 		& 	$30$		&   	$300$    	&	$26^\circ$ & 	$-0.142$ 	& 	$1.122$	&  	$5.53e-03$ 	&  	$0.329$ 	&  	$0.452$ 		& 	$32.5$	&   	$300$\\\hline
    $11^\circ$ & 	$-0.138$ 	& 	$1.194$	&  	$3.14e-04$ 	&  	$0.330$ 	&  	$0.113$ 		& 	$30$		&   	$300$    	&	$27^\circ$ & 	$-0.143$ 	& 	$1.092$	&  	$4.35e-03$ 	&       $0.329$   &      $0.410$            	&       $32.15$   &       $300$ \\\hline
    $12^\circ$ & 	$-0.126$ 	& 	$1.236$	&  	$3.06e-04$ 	&       $0.329$   &      $0.114$            	&       $29.9$      &      $300$      &	$28^\circ$ & 	$-0.144$ 	& 	$1.053$	&  	$3.76e-03$ 	&  	$0.329$ 	&  	$0.376$ 		& 	$32$		&   	$290$\\\hline
    $13^\circ$ & 	$-0.117$ 	& 	$1.272$	&  	$3.12e-04$ 	&  	$0.498$ 	&  	$0.104$ 		& 	$29.75$	&   	$300$    	&	$29^\circ$ & 	$-0.144$ 	& 	$1.015$	&  	$2.49e-03$ 	&  	$0.329$ 	&  	$0.302$ 		& 	$31.8$	&   	$285$ \\\hline
    $14^\circ$ & 	$-0.114$ 	& 	$1.286$	&  	$3.72e-04$ 	&  	$0.500$ 	&  	$0.113$ 		& 	$29.9$	&   	$305$    	&	$30^\circ$ & 	$-0.148$ 	& 	$0.983$	&  	$2.17e-03$ 	&       $0.329$   &      $0.282$            	&       $31.45$   &       $325$\\\hline
    $15^\circ$ & 	$-0.107$ 	& 	$1.289$	&  	$4.84e-04$ 	&  	$0.500$ 	&  	$0.136$ 		& 	$30$		&   	$300$    	&&&&&&&&\\\hline
  \end{tabular}
  \label{Table:CL-parametrs}
\end{table*}

\begin{table*}[t] 
  \caption{$C_D$ model parameters for AOAs $0^\circ$ to $30^\circ$. The parameter $m$ is the slope of drift function,  $C_{Do}$ the stable fix point, $\beta$ the optimized diffusion 	function, $\tilde{F}$ the correction factor for diffusion function to incorporate the model extension, $A$ the constant to fix the oscillation 		amplitude, $T$ the most dominant oscillation period and  $k_o$ half the average breathing length.}
  \center
  \scriptsize
  \begin{tabular}{| c | c | c | c | c | c | c | c || c | c | c | c | c | c | c | c |}
    \hline
    AOA 		& 	$m$   	& 	$C_{Do}$ &	$\beta$		&  	$F$		& 	$A$     		&   	$T$		&   	$k_o$	&	AOA 		& 	$m$   	& 	$C_{Do}$ &	$\beta$		&  	$F$		& 	$A$     		&   	$T$		&   	$k_o$\\\hline
    $0^\circ$ 	& 	$-0.171$ 	& 	$0.052$	&  	$1.79e-05$ 	&  	$0.33$ 	&  	$0.024$ 		& 	$16$		&   	$50$    	&	$16^\circ$ & 	$-0.162$ 	& 	$0.162$	&  	$9.22e-05$ 	&       $0.23$   	&      $0.061$         	&       $24.25$   &      $210$ \\\hline
    $1^\circ$ 	& 	$-0.190$ 	& 	$0.054$	&  	$1.67e-05$ 	&  	$0.33$ 	&  	$0.022$ 		& 	$16$		&   	$48$    	&	$17^\circ$ & 	$-0.163$ 	& 	$0.185$	&  	$1.37e-04$ 	&  	$0.23$ 	&  	$0.074$ 		& 	$24.5$	&   	$210$\\\hline
    $2^\circ$ 	& 	$-0.196$ 	& 	$0.055$	&       $1.79e-05$      &       $0.329$   	&      $0.022$         	&       $17$         &      $50$      	&	$18^\circ$ & 	$-0.172$ 	& 	$0.210$	&  	$2.22e-04$ 	&  	$0.23$ 	&  	$0.091$ 		& 	$24.75$	&   	$210$\\\hline
    $3^\circ$ 	& 	$-0.190$ 	& 	$0.056$	&       $1.90e-05$      	&       $0.33$   	&      $0.023$         	&       $17$         &      $50$      	&	$19^\circ$ & 	$-0.169$ 	& 	$0.240$	&  	$3.45e-04$ 	&       $0.23$   	&      $0.114$         	&       $25.25$   &       $210$\\\hline
    $4^\circ$ 	& 	$-0.192$ 	& 	$0.058$	&       $2.12e-05$      &       $0.33$   	&      $0.024$           	&       $17.5$      &      $100$      	&	$20^\circ$ & 	$-0.176$ 	& 	$0.270$	&  	$5.17e-04$ 	&  	$0.23$ 	&  	$0.137$ 		& 	$25.25$	&   	$300$\\\hline
    $5^\circ$ 	& 	$-0.192$ 	& 	$0.060$	&  	$2.24e-05$ 	&  	$0.33$ 	&  	$0.025$ 		& 	$18$		&   	$105$    	&	$21^\circ$ & 	$-0.169$ 	& 	$0.300$	&  	$7.03e-04$ 	&  	$0.23$ 	&  	$0.157$ 		& 	$26$		&   	$210$\\\hline
    $6^\circ$ 	& 	$-0.197$ 	& 	$0.060$	&  	$2.55e-05$ 	&       $0.33$   	&      $0.026$         	&       $18.5$      &      $90$      	&	$22^\circ$ & 	$-0.185$ 	& 	$0.336$	&  	$1.08e-03$ 	&  	$0.23$ 	&  	$0.191$ 		& 	$25.5$	&   	$210$\\\hline
    $7^\circ$ 	& 	$-0.203$ 	& 	$0.065$	&  	$2.86e-05$ 	&  	$0.23$ 	&  	$0.030$ 		& 	$23.8$	&   	$95$    	&	$23^\circ$ & 	$-0.185$ 	& 	$0.369$	&  	$1.45e-03$ 	&  	$0.23$ 	&  	$0.217$ 		& 	$25.66$	&   	$210$\\\hline	
    $8^\circ$ 	& 	$-0.200$ 	& 	$0.068$	&  	$3.20e-05$ 	&  	$0.23$ 	&  	$0.031$ 		& 	$23.85$	&   	$120$    	&	$24^\circ$ & 	$-0.191$ 	& 	$0.402$	&  	$2.01e-03$ 	&       $0.23$   	&      $0.245$            	&       $25.65$   &       $300$\\\hline
    $9^\circ$ 	& 	$-0.206$ 	& 	$0.072$	&  	$3.56e-05$ 	&       $0.23$   	&      $0.032$           	&       $23.85$    &      $120$       &	$25^\circ$ & 	$-0.193$ 	& 	$0.431$	&  	$2.38e-03$ 	&  	$0.23$ 	&  	$0.278$ 		& 	$25.65$	&   	$300$\\\hline
    $10^\circ$ & 	$-0.206$ 	& 	$0.077$	&  	$3.89e-05$ 	&  	$0.23$ 	&  	$0.034$ 		& 	$23.85$	&   	$95$    	&	$26^\circ$ & 	$-0.188$ 	& 	$0.467$	&  	$2.65e-03$ 	&  	$0.23$ 	&  	$0.302$ 		& 	$25.8$	&   	$300$ \\\hline
    $11^\circ$ & 	$-0.200$ 	& 	$0.083$	&  	$4.19e-05$ 	&  	$0.23$ 	&  	$0.036$ 		& 	$23.9$	&   	$210$    	&	$27^\circ$ & 	$-0.194$ 	& 	$0.503$	&  	$2.81e-03$ 	&       $0.23$   	&      $0.301$           	&       $25.5$   	&       $300$\\\hline
    $12^\circ$ & 	$-0.198$ 	& 	$0.092$	&  	$4.50e-05$ 	&       $0.23$   	&      $0.038$         	&       $23.9$      &      $210$       &	$28^\circ$ & 	$-0.195$ 	& 	$0.552$	&  	$2.63e-03$ 	&  	$0.23$ 	&  	$0.291$ 		& 	$25.5$	&   	$305$\\\hline
    $13^\circ$ & 	$-0.192$ 	& 	$0.108$	&  	$4.89e-05$ 	&  	$0.23$ 	&  	$0.039$ 		& 	$23.9$	&   	$210$    	&	$29^\circ$ & 	$-0.195$ 	& 	$0.590$	&  	$2.33e-03$ 	&  	$0.23$ 	&  	$0.271$ 		& 	$25.3$	&   	$250$\\\hline
    $14^\circ$ & 	$-0.184$ 	& 	$0.123$	&  	$5.92e-05$ 	&  	$0.23$ 	&  	$0.045$ 		& 	$24$		&   	$210$    	&	$30^\circ$ & 	$-0.198$ 	& 	$0.623$	&  	$2.04e-03$ 	&       $0.23$   	&      $0.255$           	&       $25.1$   	&       $325$\\\hline
    $15^\circ$ & 	$-0.175$ 	& 	$0.140$	&  	$6.89e-05$ 	&  	$0.23$ 	&  	$0.050$ 		& 	$24.15$	&   	$210$    	&&&&&&&&\\\hline 	
  \end{tabular}
  \label{Table:CD-parametrs}
\end{table*}

\section{Model integration into AeroDyn}\label{sec.Mod-int-AD}
The stochastic lift and drag model described in Section \ref{sec.Mod}
is integrated into AeroDyn to obtain aerodynamic forces with local
dynamics on the rotating blade. The model is added in form of a
    routine based on model equation (\ref{eq:Ext mod}) as an alternate
    to static airfoil data files used by the classical BEM model in
    AeroDyn. The characteristics of model related parameters given
in Tables \ref{Table:CL-parametrs} and \ref{Table:CD-parametrs} are
imported as input files to the model routine. The axial and tangential
induction factors are estimated (using mean lift and drag
coefficients) through an iterative procedure described in
Figure~\ref{fig:Flowchart-BEM}.
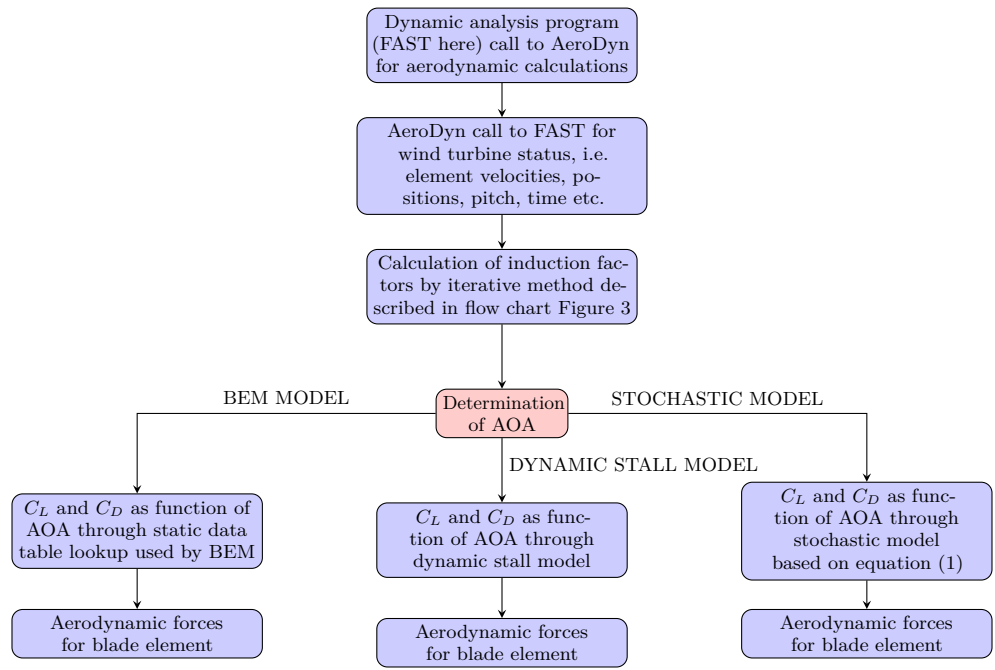
\begin{figure*}[t]
  \center
  
  \begin{tikzpicture}[scale=0.8, node distance = 2cm, auto,transform shape]
    
    \tikzstyle{block} = [rectangle, draw, fill=blue!20,text width=13em, text centered, rounded corners, minimum height=2em]
    \node [block] (init) {Dynamic analysis program (FAST here) call to AeroDyn for aerodynamic calculations};
    
    \tikzstyle{block} = [rectangle, draw, fill=blue!20, text width=14em, text centered, rounded corners, minimum height=2em]
    \node [block, below of=init,node distance = 2cm] (identify) {AeroDyn call to FAST for wind turbine status, i.e. element velocities, positions, pitch, time 	etc.};
    
    \tikzstyle{block} = [rectangle, draw, fill=blue!20, text width=13em, text centered, rounded corners, minimum height=2em]
    \node [block, below of=identify,node distance = 2cm] (evaluate) {Calculation of induction factors by iterative method described in flow chart Figure \ref{fig:Flowchart-BEM}};
    
    \tikzstyle{block} = [rectangle, draw, fill=red!20, text width=6em, text centered, rounded corners, minimum height=2em]
    \node [block, below of=evaluate,node distance = 2.1cm] (decide) {Determination of AOA};
    
    \tikzstyle{block} = [rectangle, draw, fill=blue!20, text width=12em, text centered, rounded corners, minimum height=2em]
    \node [block, left of=decide,node distance=6cm,yshift=-6em] (CLCD-static) {$C_L$ and $C_D$ as function of AOA through static data table lookup used by BEM};
    
    \tikzstyle{block} = [rectangle, draw, fill=blue!20, text width=12em, text centered, rounded corners, minimum height=2em]
    \node [block, right of=decide, node distance=6cm,yshift=-6em] (CLCD-dynamic) {$C_L$ and $C_D$ as function of AOA through stochastic model based on equation (\ref{eq:Ext mod})};
    
    \tikzstyle{block} = [rectangle, draw, fill=blue!20, text width=12em, text centered, rounded corners, minimum height=2em]
    \node [block, below of=CLCD-static, node distance=1.7cm] (CNCT-static) {Aerodynamic forces for blade element};
    
    \tikzstyle{block} = [rectangle, draw, fill=blue!20, text width=12em, text centered, rounded corners, minimum height=2em]
    \node [block, below of=CLCD-dynamic, node distance=1.7cm] (CNCT-dynamic) {Aerodynamic forces for blade element};
    
    \tikzstyle{block} = [rectangle, draw, fill=blue!20, text width=12em, text centered, rounded corners, minimum height=2em]
    \node [block, below of=decide, node distance=2.1cm] (CLCD-Dstall) {$C_L$ and $C_D$ as function of AOA through dynamic stall model};	
    
    \tikzstyle{block} = [rectangle, draw, fill=blue!20, text width=12em, text centered, rounded corners, minimum height=2em]
    \node [block, below of=CLCD-Dstall, node distance=1.7cm] (CNCT-Dstall) {Aerodynamic forces for blade element};

    \tikzstyle{line} = [draw, -stealth]
    \path [line] (init) -- (identify);
    \path [line] (identify) -- (evaluate);
    \path [line] (evaluate) -- (decide);
    
    \path [line] (decide) -| node[yshift=1.5em,near start] {BEM MODEL} (CLCD-static);
    \path [line] (decide) -| node[yshift=0.1em,near start]{STOCHASTIC MODEL} (CLCD-dynamic);
    \path [line] (decide) -- node[yshift=-0.5em,near start]{DYNAMIC STALL MODEL} (CLCD-Dstall);
    
    \path [line] (CLCD-static) -- (CNCT-static);
    \path [line] (CLCD-dynamic) -- (CNCT-dynamic);
    \path [line] (CLCD-Dstall) -- (CNCT-Dstall);
    
  \end{tikzpicture}
  \caption{Flow chart for aerodynamic calculations.}
  \label{fig:Flowchart-AeroDyn}
\end{figure*}

\subsection{Numerical setup}\label{subsec-Num-setup}
Before starting the aerodynamic computations, the AeroDyn input files
described in Section~\ref{subsec-AeroDyn-input} are set up for
intended options of computations. In primary file the wind reference
height, air density and air kinematic viscosity are taken as 42.7 m,
1.225 kg/m$^3$ and 1.464e$^{-5}$ m$^2$/s, respectively.

The aerodynamic calculations are performed for a three-bladed rotor
having radius of 13.76\,m. The blade is divided into 10 equal segments
along the radius. The local design specifications for each element are
given in Table \ref{Table:ElmSpec} containing the element nodal radius
(from blade hub centre to the centre of element), twist angle, span
and chord length. All three blades have the same distribution yielding
identical aerodynamics for same pitch angle -1$^\circ$. The elemental
pitch can be obtained by adding the blade pitch angle to the element's local
twist angle. The computations are carried out for all 10 elements;
however, the results here are presented for element number 5 only to
avoid the repetition of similar statistics.
\begin{table}
  \caption{Blade local design parameters along the span. Taken from FAST archive (Test03\_AD.ipt).}
  \center
  \scriptsize\addtolength{\tabcolsep}{5pt}
  \begin{tabular}{| c | c | c | c | c |}
    \hline
    No. of 		&  	Nodal 		&	Twist 		& 	Element 			&	Chord 	\\
    element		& 	radius [m]  	& 	angle [$^\circ$] &       span [m]                     &       length [m]                   \\\hline
    $1$ 			&	$1.81$		&  $5.80$ 	   		& 	$1.26$			&  	$0.86$ 			\\\hline
    $2$ 			&	$3.07$		&  $5.20$ 	   		& 	$1.26$			&  	$1.05$ 			\\\hline
    $3$ 			&	$4.33$		&  $4.66$ 	   		& 	$1.26$			&       $1.15$      		\\\hline
    $4$ 			&	$5.58$		&  $3.73$ 	   		& 	$1.26$			&       $1.12$      	  	\\\hline
    $5$ 			&	$6.84$		&  $2.64$ 	   		& 	$1.26$			&  	$1.05$ 			\\\hline
    $6$ 			&	$8.10$		&  $1.59$ 	   		& 	$1.26$			&  	$0.98$ 			\\\hline
    $7$ 			&	$9.36$		&  $0.73$ 	   		& 	$1.26$			&       $0.89$      		\\\hline
    $8$ 			&	$10.61$		&  $0.23$ 	   		& 	$1.26$			&       $0.78$      	  	\\\hline
    $9$ 			&	$11.87$		&  $0.08$ 	   		& 	$1.26$			&  	$0.65$ 			\\\hline
    $10$ 		&	$13.13$		&  $0.03$ 	   		& 	$1.26$			&  	$0.49$ 			\\\hline
  \end{tabular}
  \label{Table:ElmSpec}
\end{table}

Since the model is based on aerodynamic characteristics of an FX
79-W-151A, for comparison purpose the airfoil data file of AeroDyn is
replaced with measured aerodynamic characteristics of this airfoil
(for measurement details see~\cite{Luhur14}).

For wind input, the full field turbulence simulated binary wind data
file is used. The file represents all three components of the wind
vector created with TurbSim program. The components are variable in
time and space obtained at each element each moment by subroutine
interpolations. A summary of the meteorological parameters of the
wind data file is given in Table~\ref{Table:FF-wind}.
\begin{table}
  \caption{Meteorological  boundary conditions of the wind data file. Taken from FAST archive.} 
  \center
  \scriptsize\addtolength{\tabcolsep}{0pt}
  \begin{tabular}{p{0.6\linewidth}cp{0.35\linewidth}}
    \hline
    Turbulence model used							&:& IEC von Karman\\				
    IEC standard								         &:& IEC 61400-1 Ed. 2: 1999\\
    Turbulence characteristic						         &:& A\\		
    IEC turbulence type							         &:& Normal turbulence model\\		
    Reference height (hub-height)					         &:& 42.7 m\\
    Reference wind speed (mean speed at hub-height)	         &:& 12 m/s\\
    Power law exponent 						         &:& 0.2\\
    Surface roughness length					                  &:& 0.03\\ 
    Interpolated hub-height turbulence intensity		         &:& 15\%\\
    \hline			
  \end{tabular}
  \label{Table:FF-wind}
\end{table}

It has to be stressed that the model parameters have been derived from
wind tunnel measurements in fractal square grid generated stationary
turbulence, while the simulation uses synthetic turbulent inflow
according to Table~\ref{Table:FF-wind}. To compensate at least
partially for the different flow situations, the diffusion function of
the basic model equation (\ref{eq:Langevin}) is multiplied with a
correction factor obtained by dividing synthetic wind input
interpolated hub-height turbulence intensity with fractal square grid
generated turbulence intensity. Due to the high complexity of
turbulence, this can nevertheless not ensure completely comparable
flow situations.

As described in Section \ref{sec:AD}, AeroDyn can not be operated
independently, but has to be initiated by a dynamic analysis code.
Here it is invoked by FAST, which uses a combined modal and multi-body
dynamics representation. In FAST, the wind turbine blades and tower
are modeled by applying the linear representation considering small
deflections with mode shapes (degrees of freedom (DOF)) listed in
Table \ref{Table:FAST-parameters} \cite{Jonkman05,Cordle10}. FAST
collects the basic information from its primary input file containing
the details of wind turbine operating conditions and the basic
geometry. For additional information such as blade properties, tower
properties, furling properties, wind time histories and the
aerodynamic characteristics, FAST reads some supplementary files
\cite{Jonkman05}.
\begin{table}
  \caption{Wind turbine mode shapes and configuration. Taken from FAST archive (Test03.fst).} 
  \center
  \scriptsize\addtolength{\tabcolsep}{0pt}
  \begin{tabular}{p{0.80\linewidth}cp{0.15\linewidth}}
    \hline
    Number of blades used								&:& 3\\				
    First and second flap-wise blade mode DOF				&:& Yes\\		
    First edge-wise blade mode DOF				         		&:& Yes\\		
    Drivetrain rotational-flexibility DOF				         		&:& Yes\\
    Generator DOF								         		&:& Yes\\
    Yaw DOF									                   &:& Yes\\
    First and second fore-aft tower bending-mode DOF	         		&:& Yes\\
    First and second side-to-side tower bending-mode DOF  		&:& Yes\\ 
    Initial or fixed rotor speed						         		&:& 53.33 rpm\\
    Blade tip radius from rotor apex 	         					&:& 13.76 m\\
    Hub radius from rotor apex to blade root			         		&:& 1.18 m\\
    Tower height from ground level to rotor-shaft (hub-height)	&:& 42.7 m\\
    \hline			
  \end{tabular}
  \label{Table:FAST-parameters}
\end{table}
 
Once FAST calls AeroDyn and exchanges the information on the model
status including elemental pitch and velocities, AeroDyn starts to
compute the elemental aerodynamic forces. The velocity components are
expressed normal and tangential to the plane of rotation.

\subsection{Results}\label{subsec-Results} 
\begin{figure*}[t]
  \center
  \small
  \includegraphics[width=0.85\textwidth]{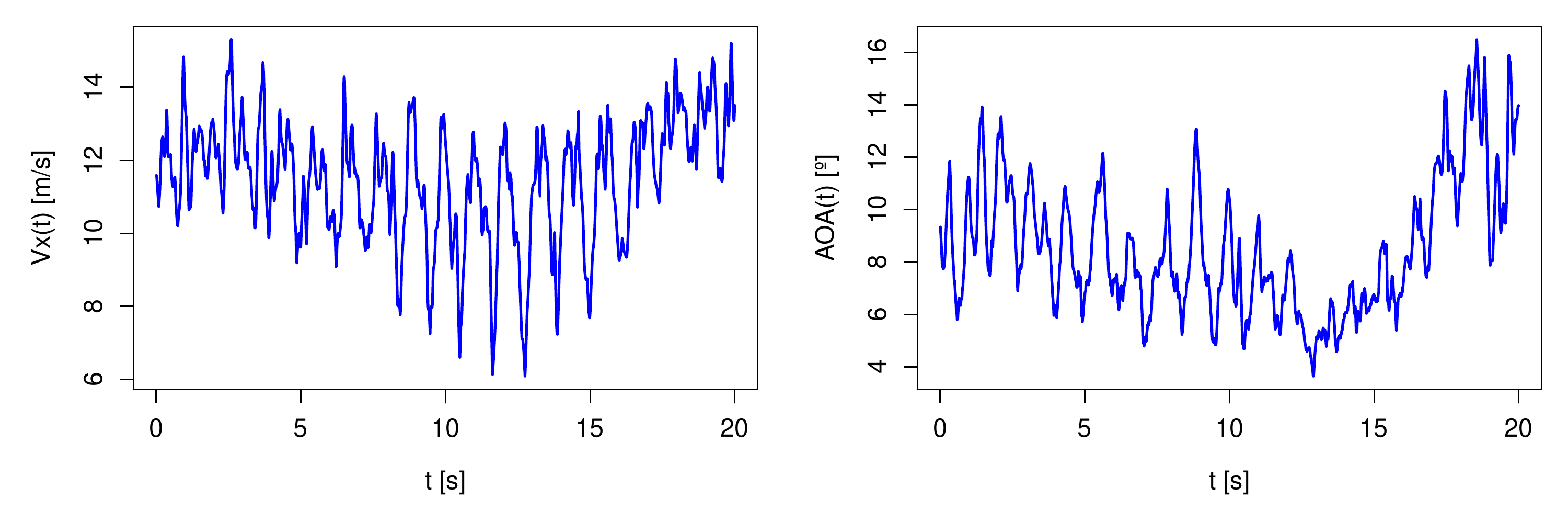}
  \put(-303,122){(a)}	
  \put(-95,122){(b)}	
  \caption{Excerpt of local axial wind velocity component and AOA time
    series for a blade element. (a) Local axial wind component
    experienced by the blade element and (b) local AOA. Note the rotor
    oscillation at $T=1.13$\,s in (a) and (b), which possibly stems
    from ground boundary layer shear effects\protect\footnotemark[2].
    The oscillation in (b) is less visible because of short excerpt;
    however, it is present at the same period.}
  \label{fig:VxAOA}
\end{figure*}

Results are achieved for a blade element following the numerical setup
described in Section \ref{subsec-Num-setup}. The force calculations
are performed using the classical BEM, dynamic stall and stochastic
models according to procedure given in flow chart
Figure~\ref{fig:Flowchart-AeroDyn}. The TurbSim generated synthetic
wind is used as an input to AeroDyn. Simulations are performed for 10
realizations of 10 minutes of wind input. The synthetic wind,
containing irregular speed and irregular turbulence level, led to
strong fluctuations in local wind components and AOA as expected; see
Figure~\ref{fig:VxAOA}. The figure portrays the complex fluctuations
of the axial velocity component experienced by the blade element and
its effect on AOA dynamics; compare~\cite{Stoevesandt09}. The
variations in velocity component are proportional to the AOA, i.e.,
the higher the wind fluctuations, the higher the AOA variation. A
pronounced oscillation at $T=1.13$\,s is visible in
Figure~\ref{fig:VxAOA}, which possibly stems from boundary layer shear
effects, as the tower-blade interaction was not included in the
simulations \footnote{%
  The appearance of harmonics of the 1P period seems to be typical for
  the rotating frame of reference of the rotor \cite{Jonkman2014}.%
}.
 	
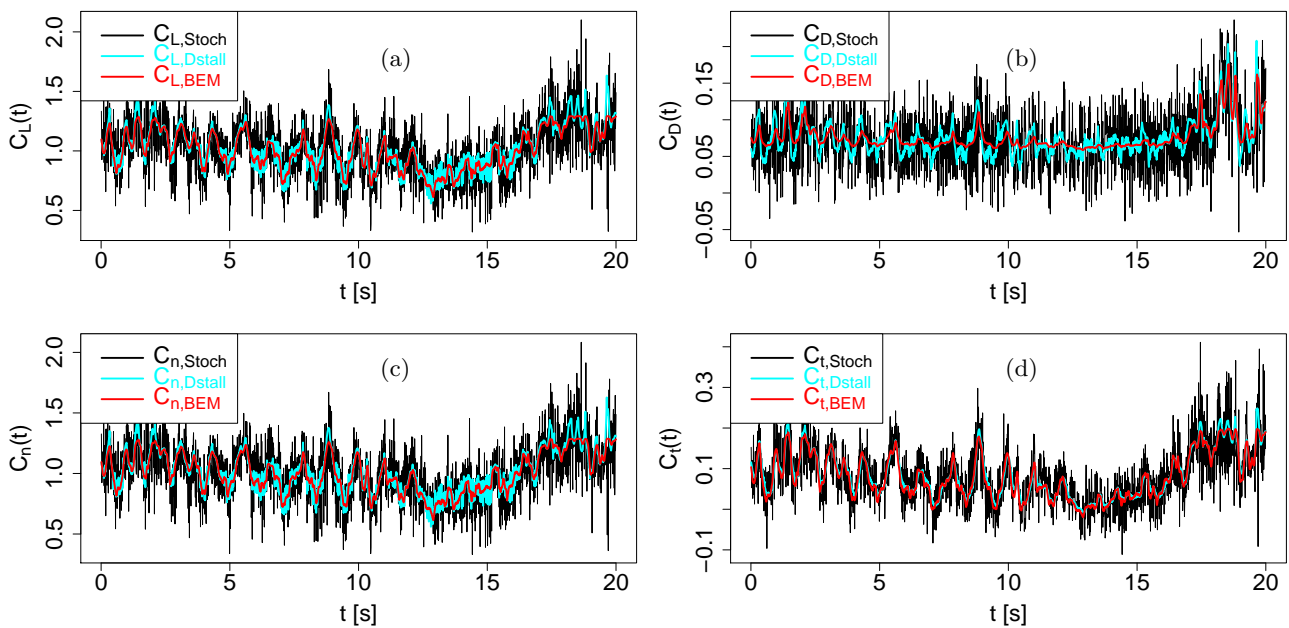
\begin{figure*}[t]
  \center
  \small
  \includegraphics[width=0.95\textwidth]{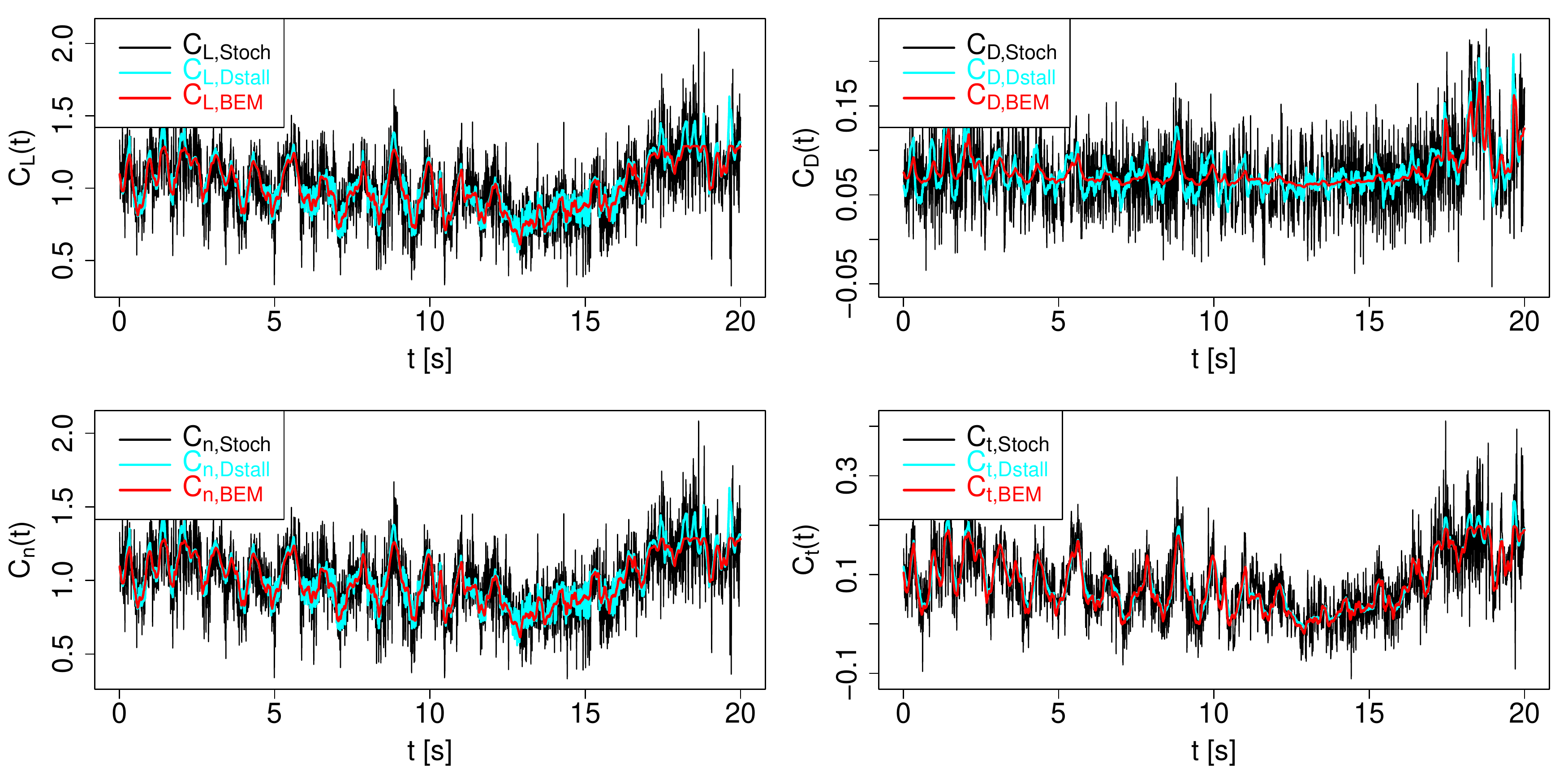}
  \put(-345,217){(a)}	
  \put(-111,217){(b)}	
  \put(-345,100){(c)}	
  \put(-111,100){(d)}	
  \caption{Excerpt of the stochastic model, the dynamic stall
      model and the classical BEM model aerodynamic forces time series
      for a blade element. Black represents the stochastic model
      forces, cyan the dynamic stall model forces and red the
      classical BEM model forces (color online). (a) $C_{L,Stoch}(t)$,
      $C_{L,Dstall}(t)$ and $C_{L,BEM}(t)$, (b) $C_{D,Stoch}(t)$,
      $C_{D,Dstall}(t)$ and $C_{D,BEM}(t)$, (c) $C_{n,Stoch}(t)$,
      $C_{n,Dstall}(t)$ and $C_{n,BEM}(t)$, and (d) $C_{t,Stoch}(t)$,
      $C_{t,Dstall}(t)$ and $C_{t,BEM}(t)$.}
  \label{fig:TSeries}
\end{figure*}

The blade aerodynamics is function of the AOA, therefore, a change
in AOA means a change in aerodynamic forces; compare
Figure~\ref{fig:VxAOA}(b) with Figure~\ref{fig:TSeries}.
Figure~\ref{fig:TSeries} shows the resulting aerodynamic forces
behavior for the selected element, where the force coefficient signals
obtained with classical BEM represent the mean force dynamics as
expected, being dependent on mean aerodynamic characteristics of the
airfoil. The force coefficient signals achieved with the dynamic stall
model represent the force dynamics with small fluctuations around the
mean, while the force coefficient signals contributed by the
stochastic model represent the force dynamics with extended local
dynamics around the mean compared to the dynamic stall model. The
means of the dynamic stall and stochastic models' local force dynamics
along the signals match almost perfectly with the classical BEM model
force signals.

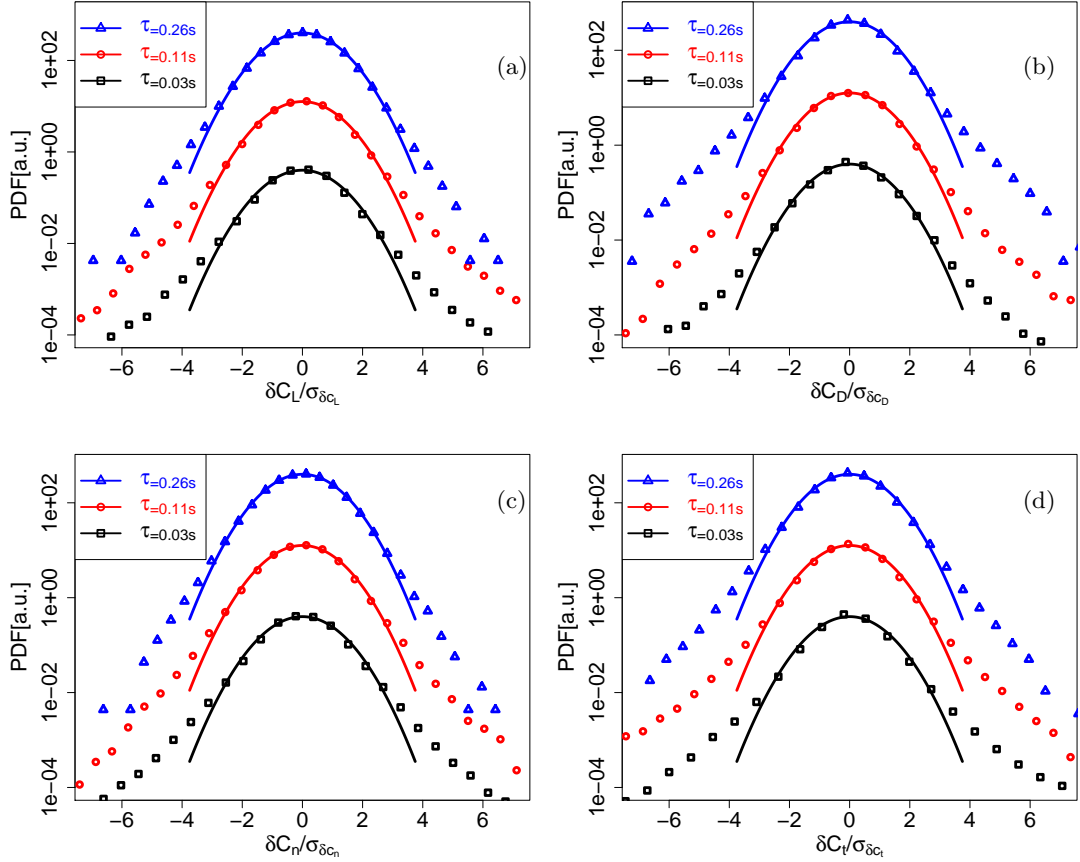
\begin{figure*}[t]
  \center
  \small
  \includegraphics[width=0.8\textwidth]{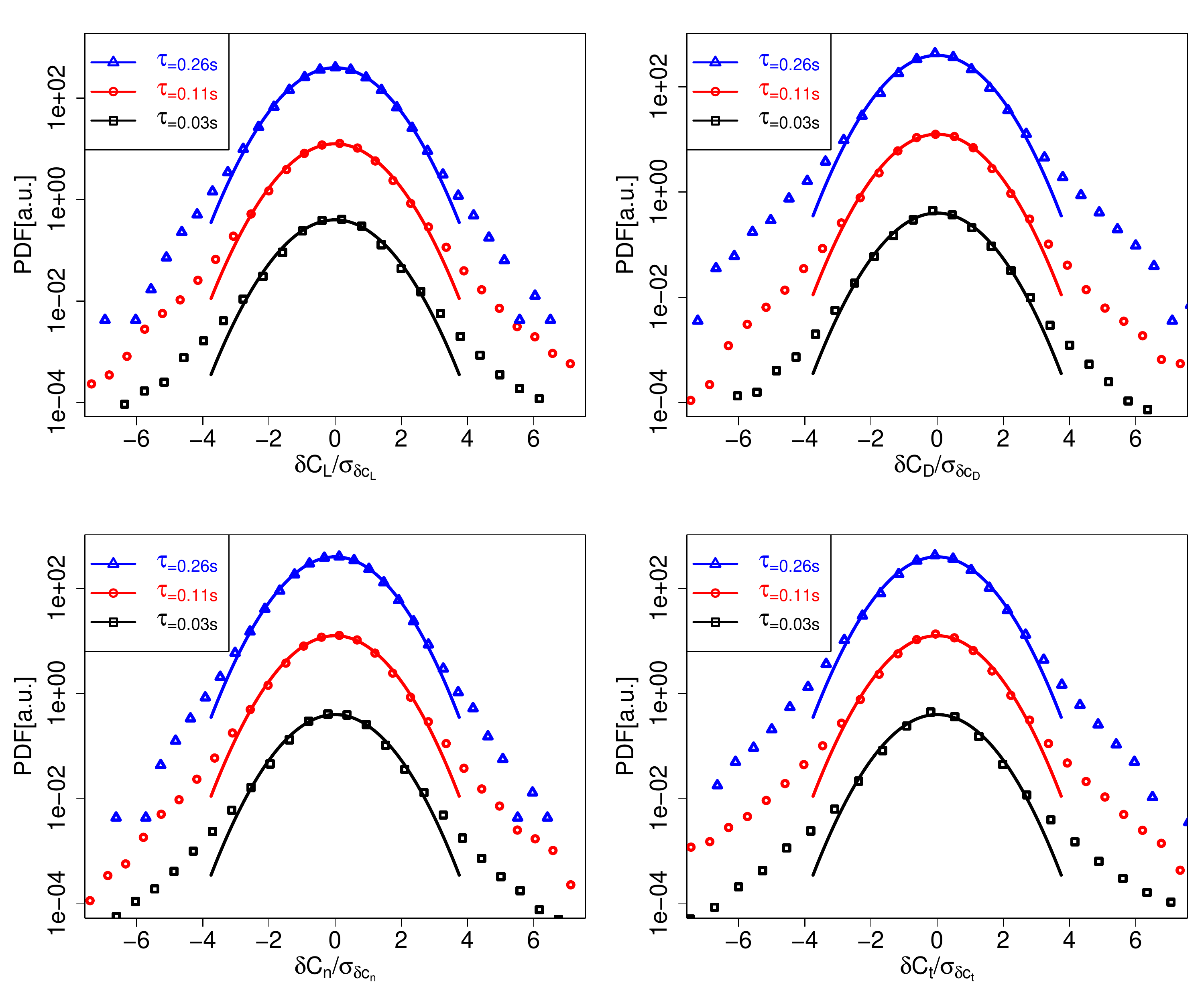}
  \put(-223,303){(a)}		
  \put(-26,303){(b)}			
  \put(-223,139){(c)}		
  \put(-26,139){(d)}		
  \caption{Increment PDFs of the stochastic model force coefficients
    for a blade element at time lags $\tau=(0.03,0.11,0.26)$\,s in
    ascending order from bottom to top. The PDFs are added with a
    Gaussian fit having identical standard deviation (solid line) and
    shifted vertically for clarity of the display. The force
    coefficients are normalized with their standard deviations. (a)
    Lift coefficient increment $\delta{C_L}(t,\tau)$ PDFs, (b) Drag
    coefficient increment $\delta{C_D}(t,\tau)$ PDFs, (c) Normal force
    coefficient increment $\delta{C_n}(t,\tau)$ PDFs, and (d)
    Tangential force coefficient increment $\delta{C_t}(t,\tau)$
    PDFs.}
  \label{fig:IncPDFs-model}
\end{figure*}

\begin{figure*}[t]
  \center
  \small
  \includegraphics[width=0.8\textwidth]{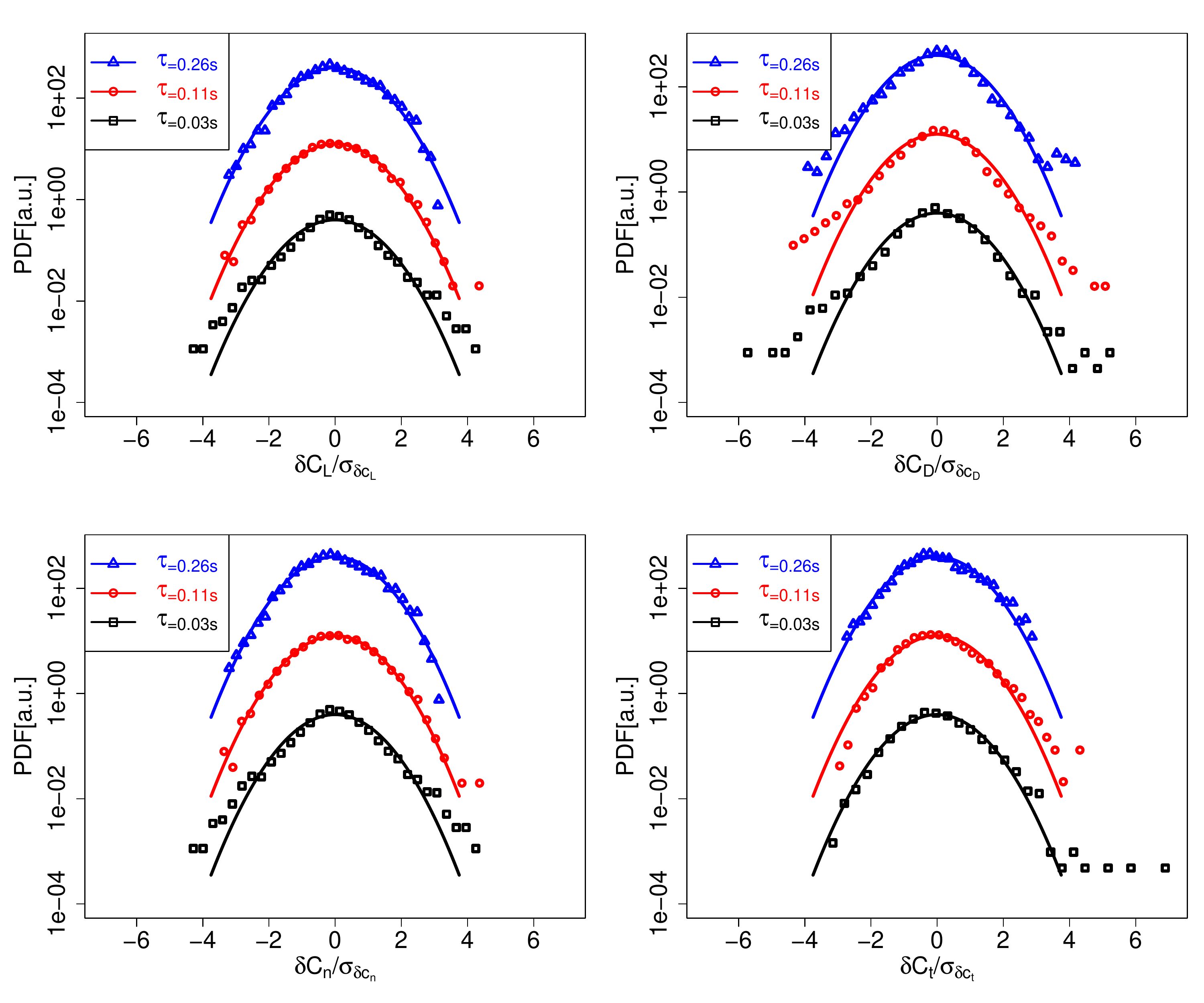}
  \put(-223,303){(a)}		
  \put(-26,303){(b)}			
  \put(-223,139){(c)}		
  \put(-26,139){(d)}		
  \caption{Increment PDFs of the dynamic stall model force
    coefficients for a blade element at time lags
    $\tau=(0.03,0.11,0.26)$\,s in ascending order from bottom to top.
    The PDFs are added with a Gaussian fit having identical standard
    deviation (solid line) and shifted vertically for clarity of the
    display. The force coefficients are normalized with their standard
    deviations. (a) Lift coefficient increment $\delta{C_L}(t,\tau)$
    PDFs, (b) Drag coefficient increment $\delta{C_D}(t,\tau)$ PDFs,
    (c) Normal force coefficient increment $\delta{C_n}(t,\tau)$ PDFs,
    and (d) Tangential force coefficient increment
    $\delta{C_t}(t,\tau)$ PDFs.}
  \label{fig:IncPDFs-Dstall}
\end{figure*}

\begin{figure*}[t]
  \center
  \small
  \includegraphics[width=0.8\textwidth]{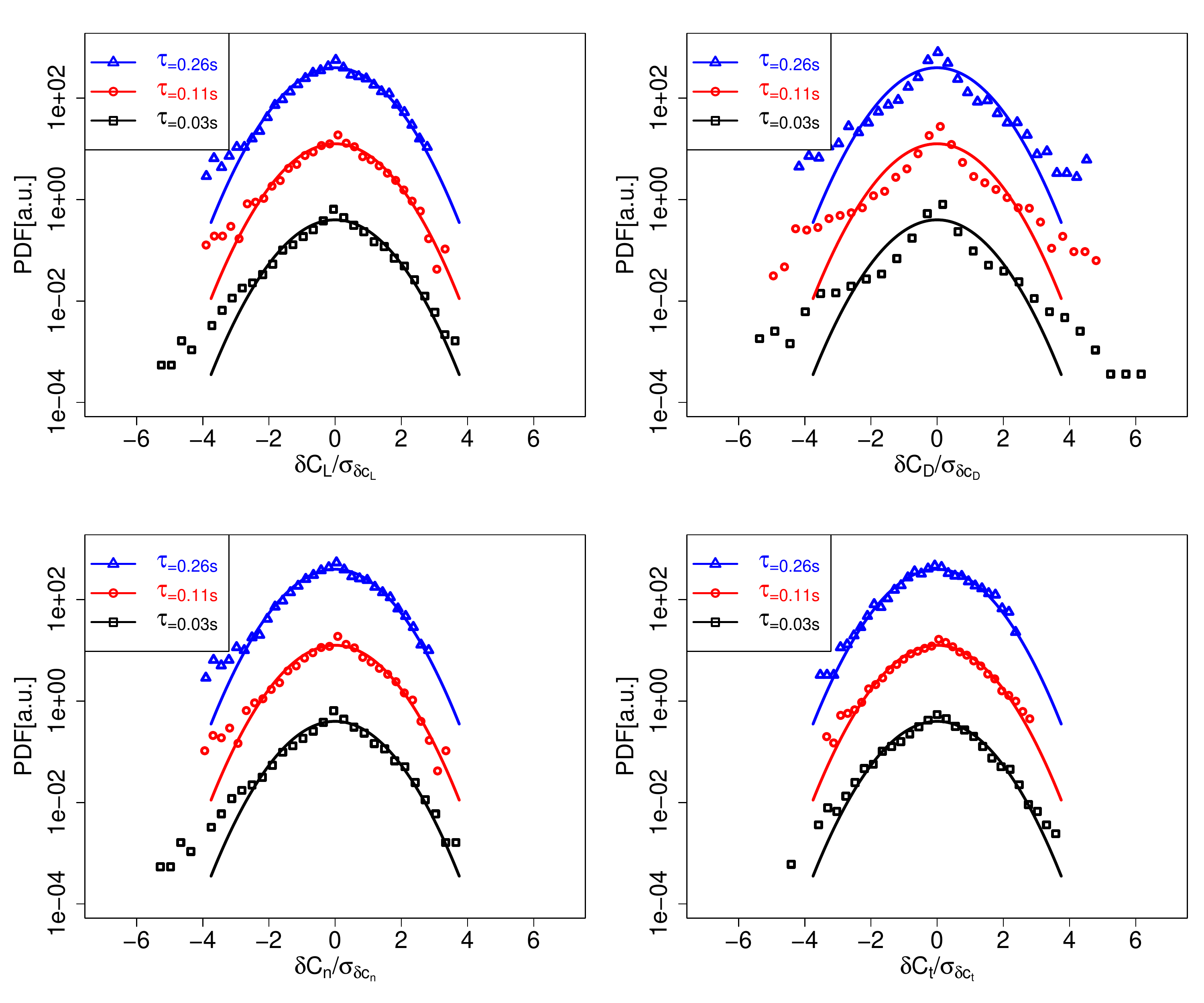}
  \put(-223,303){(a)}		
  \put(-26,303){(b)}			
  \put(-223,139){(c)}		
  \put(-26,139){(d)}	
  \caption{Increment PDFs of the classical BEM model force
    coefficients for a blade element at time lags
    $\tau=(0.03,0.11,0.26)$\,s in ascending order from bottom to top.
    The PDFs are added with a Gaussian fit having identical standard
    deviation (solid line) and shifted vertically for clarity of the
    display. The force coefficients are normalized with their standard
    deviations. (a) Lift coefficient increment $\delta{C_L}(t,\tau)$
    PDFs, (b) Drag coefficient increment $\delta{C_D}(t,\tau)$ PDFs,
    (c) Normal force coefficient increment $\delta{C_n}(t,\tau)$ PDFs,
    and (d) Tangential force coefficient increment
    $\delta{C_t}(t,\tau)$ PDFs.}
  \label{fig:IncPDFs-static}
\end{figure*}

In order to investigate the quality of the stochastic model results, their
statistical properties are compared with the classical BEM and
    dynamic stall models' results in terms of increment probability
density functions (PDFs). The increments in our case are the
differences of an aerodynamic force over a specific time lag described
as \cite{Morales12}
\begin{equation}
  \delta{x}(t,\tau)=x(t+\tau)-x(t).
  \label{eq:inrement}
\end{equation}
The increment statistics are basically two-point statistics, which
determine the nature of a parameter variation against a selected time
lag. Using equation (\ref{eq:inrement}), the increment signals of the
aerodynamic forces are derived for different time lags. Increment
PDFs of the stochastic, dynamic stall and classical BEM models'
results for different time lags are shown in Figures
\ref{fig:IncPDFs-model}, \ref{fig:IncPDFs-Dstall} and
\ref{fig:IncPDFs-static}, respectively.
					
For quantitative comparison, further kurtosis and standard deviations
of the increment signals have been estimated. The kurtosis is
calculated using the expression
\begin{equation}
  \gamma_{_2}=\frac{ \langle(x-\bar{x})^4\rangle}{\sigma_{x}^4}-3,
  \label{eq:Kurtosis}
\end{equation}
where $\gamma_{_2}$ is the excess kurtosis,
$\langle(x-\bar{x})^4\rangle$ the fourth moment around the mean and
$\sigma_{x}^4$ the square of the variance of the probability
distribution. The $\gamma_{_2}=0$ resemble a normal distribution,
$\gamma_{_2}>0$ a distribution with sharp peak and long heavy tails,
and $\gamma_{_2}<0$ a distribution with round peak and short light
tails. The excess kurtosis is a measure of the deviation of PDF
shape from the normal distribution.

\begin{table}[h!] 
  \caption{Kurtosis and standard deviation of the increment signals of the stochastic model force coefficients at different time lags.} 
  \center
  \scriptsize\addtolength{\tabcolsep}{0.3pt}
  \begin{tabular}{| c | c | c | c | c | c | c | c |c |}
    \hline
    \multicolumn{1}{|c|}{Inrement } &  \multicolumn{4}{c|}{Kurtosis} &  \multicolumn{4}{c|}{Standard deviation}\\
    \hline
    $\tau$[s] 	&  $\delta{C_L}$  & 	$\delta{C_D}$ 	&	$\delta{C_n}$	&  $\delta{C_t}$	& 	$\delta{C_L}$   & $\delta{C_D}$ 	&	$\delta{C_n}$	&  $\delta{C_t}$\\\hline
    $0.03$ 	&  $1.80$ 	   	& 	$1.37$		&  	$1.66$ 		&  	$3.04$ 		&  	$0.26$ 		& $0.06$			&   $0.26$  		& $0.08$		\\\hline
    $0.11$ 	&  $0.98$ 	   	& 	$1.01$		&  	$0.90$ 		&  	$1.59$ 		&  	$0.29$ 		& $0.06$			&   $0.28$  		&  $0.09$		\\\hline
    $0.26$ 	&  $0.44$ 	   	& 	$1.42$		&       $0.40$      	&       $0.96$   		&       $0.31$         	& $0.06$     		&   $0.30$  		&  $0.09$		\\\hline
  \end{tabular}
  \label{Table:Mod-kurtosis-sd}
\end{table}

\begin{table}[h!] 
  \caption{Kurtosis and standard deviation of the increment signals of the dynamic stall model force coefficients at different time lags.} 
  \center
  \scriptsize\addtolength{\tabcolsep}{-0.7pt}
  \begin{tabular}{| c | c | c | c | c | c | c | c |c |}
    \hline
    \multicolumn{1}{|c|}{Inrement } &  \multicolumn{4}{c|}{Kurtosis} &  \multicolumn{4}{c|}{Standard deviation}\\
    \hline
    $\tau$[s] 	&  $\delta{C_L}$  & 	$\delta{C_D}$ 	&	$\delta{C_n}$	&  $\delta{C_t}$	& 	$\delta{C_L}$   & $\delta{C_D}$ 	&	$\delta{C_n}$	&  $\delta{C_t}$\\\hline
    $0.03$ 	&  $1.38$ 	   	& 	$2.23$		&  	$1.40$ 		&  	$1.01$ 		&  	$0.06$ 		& $0.01$			&   $0.06$  		& $0.01$		\\\hline
    $0.11$ 	&  $0.06$ 	   	& 	$1.84$		&  	$0.07$ 		&  	$0.33$ 		&  	$0.11$ 		& $0.03$			&   $0.11$  		&  $0.03$		\\\hline
    $0.26$ 	&  $-0.09$    	& 	$1.34$		&       $-0.09$      	&       $-0.05$   		&       $0.17$         	& $0.04$     		&   $0.17$  		&  $0.05$		\\\hline
  \end{tabular}
  \label{Table:Dstall-kurtosis-sd}
\end{table}

\begin{table}[h!] 
  \caption{Kurtosis and standard deviation of the increment signals of the classical BEM model force coefficients at different time lags.} 
  \center
  \scriptsize\addtolength{\tabcolsep}{0.0pt}
  \begin{tabular}{| c | c | c | c | c | c | c | c |c |}
    \hline
    \multicolumn{1}{|c|}{Inrement } &  \multicolumn{4}{c|}{Kurtosis} &  \multicolumn{4}{c|}{Standard deviation}\\
    \hline
    $\tau$[s] 	&  $\delta{C_L}$  & 	$\delta{C_D}$ 	&	$\delta{C_n}$	&  $\delta{C_t}$ & 	$\delta{C_L}$   & $\delta{C_D}$ &	$\delta{C_n}$	&  $\delta{C_t}$\\\hline
    $0.03$ 	&  $1.37$ 	   	& 	$6.62$		&  	$1.42$ 		&  	$0.72$ 	 &  	$0.03$ 		& $0.01$		   &  $0.03$ 		&  $0.01$    	\\\hline
    $0.11$ 	&  $0.78$ 	   	& 	$4.43$		&  	$0.81$ 		&  	$0.28$ 	 &  	$0.08$		&  $0.02$	   	   &  $0.08$		&  $0.03$    	\\\hline
    $0.26$ 	&  $0.49$ 	   	& 	$3.31$		&       $0.51$      	&       $-0.02$   	 &      $0.14$ 		&  $0.03$  	   &  $0.14$ 		&  $0.05$      	\\\hline
  \end{tabular}
  \label{Table:Static-kurtosis-sd}
\end{table}

Figure \ref{fig:IncPDFs-model} presents the increment PDFs for
stochastic model results, where all four aerodynamic force
coefficients look similar at all three time lags except some
differences at the tails. All PDFs almost resemble the normal
distribution shape up to $\pm3\sigma$; compare with added Gaussian
fit. At the tails, the PDFs deviate from the Gaussian distribution and
show intermittent behavior. To quantify for these shapes, the kurtosis
and standard deviations of the increment signals have been estimated
as given in Table~\ref{Table:Mod-kurtosis-sd}. The estimations
    show slight to pronounced higher positive kurtosis for all four
    aerodynamic force coefficient signals at the selected time lags,
    meaning that the increment PDFs have slight to pronounced heavier
    tails. The main reason for this effect seem to be the non-linear lift and drag
characteristics of the airfoil over AOA. 
The standard deviations of the increment signals in each force
coefficient case at all three time lags are close to each other
    in magnitude; see Table \ref{Table:Mod-kurtosis-sd}.

Figure \ref{fig:IncPDFs-Dstall} presents the increment PDFs for
    dynamic stall model results. In this case, the increment PDFs
    except those of the drag coefficient, resemble similar shape with
    minor differences at the tails. The drag increment PDFs present
    heavier tails compared to other three force coefficients. However,
    the increment PDFs of all forces correspond fairly to the
    normal distribution up to $\pm3\sigma$ (compare with added
    Gaussian fit) at all three time lags. The estimated kurtosis
    values given in Table \ref{Table:Dstall-kurtosis-sd} indicate very
    slight to pronounced heavier tails for all force coefficient
    signals except almost no deviations in case of the lift, normal force
    and tangential force coefficients at higher time lags. The
    contribution of pronounced intermittency in the force signals is
    believed to stem from the same phenomenon as described in
    stochastic model case. The standard deviations of the all three
    increment signals in each force case depict increase with an increase
    in time lag; see Table~\ref{Table:Dstall-kurtosis-sd}.
	
Similarly, Figure \ref{fig:IncPDFs-static} shows the increment PDFs
for classical BEM model results. Here, except drag coefficient, the
increment PDFs of other three force coefficients portray similar shape
for same time lags. The increment PDFs of the lift, normal force and
tangential force coefficients, resemble fairly the normal distribution
shape up to $\pm3\sigma$ (compare with added Gaussian fit), whereas
the drag coefficient correspond to the intermittent shape. The
evaluated kurtosis given in Table \ref{Table:Static-kurtosis-sd}
indicate strong heavier tails for drag coefficient increment
PDFs at all three time lags. The other three force coefficients show
slight to pronounced heavier tails for all three time lags
except negligible lighter tails for higher time lag in case of the
tangential force coefficient. The contribution of strong
intermittency in the drag coefficient as well as slight to
pronounced intermittency in other force coefficients can
    probably be addressed to the same phenomenon as in the stochastic
    model case. The standard deviations of the force coefficient
    signals follow the same behavior as in the dynamic stall model
    case above; compare Tables \ref{Table:Dstall-kurtosis-sd} and
    \ref{Table:Static-kurtosis-sd}. 

\section{Discussion and outlook}\label{sec.DisOutlook}
The stochastic model of the lift and drag dynamics for an airfoil FX
79-W-151A described in Section \ref{sec.Mod} has been integrated to
AeroDyn in the context of BEM wake model. The forces are obtained
    through the classical BEM, dynamic stall and stochastic models
    used by AeroDyn. For wind input, the full field turbulence
simulated binary wind data is used. The forces are estimated for 10
blade elements; however, the results are analyzed here for one element
only to avoid the repetition of similar statistics. With this ansatz
it could be shown that local, short-time fluctuations of aerodynamic
forces can be integrated as a stochastic model into a current WEC
simulation tool.

The comparison of force time series given in Figure
    \ref{fig:TSeries} show that the classical BEM model force
    coefficient signals represent the mean dynamics as expected, being
    dependent on mean aerodynamic characteristics of the airfoil. The
    dynamic stall model forces reflect small fluctuations around the
    mean, whereas the stochastic model introduces additional extended
    local force dynamics around the mean. The means of the local force
    dynamics achieved with dynamic stall and stochastic models seem to
    coincide with the classical BEM model force coefficient signals.
    Nevertheless, the stochastic model contributes additional extended
    dynamic response in terms of local force fluctuations, which is
    neglected by the classical BEM model fully and partially by the
    dynamic stall model.

The statistics of the stochastic and dynamic stall models
    presented in Figures \ref{fig:IncPDFs-model} and
    \ref{fig:IncPDFs-Dstall} show good consistency. The increment PDFs
    of stochastic model force coefficients resemble normal
    distributions up to $\pm3\sigma$. Similarly, the increment PDFs of
    dynamic stall model force coefficients also resemble fairly a
    normal distribution up to $\pm3\sigma$. Both the stochastic and
    dynamic stall models' force increment PDFs yield slight to
    pronounced intermittency due to the non-linear lift and drag
    characteristics of the airfoil over AOA. The magnitude of the standard
    deviations of the stochastic model force increment signals is
    significantly higher that of the dynamic stall model; compare
    Tables \ref{Table:Mod-kurtosis-sd} and
    \ref{Table:Dstall-kurtosis-sd}. That is because the dynamic stall
    model forces possess smaller magnitudes of local fluctuations
    compared to the stochastic model. Moreover, the behavior of the
    standard deviations of the increment signals is different in both
    the stochastic and dynamic stall model cases. The standard
    deviations in the stochastic model case seem nearly stable for all
    time lags, whereas for the dynamic stall model case they are
    proportional to time lags.

Similarly, the comparison of statistics between stochastic
    and classical BEM models given in Figures \ref{fig:IncPDFs-model}
    and \ref{fig:IncPDFs-static} also reflect good consistency except
    for the drag coefficient. The increment PDFs of the stochastic and
    classical BEM models' force coefficients resemble normal
    distributions up to $\pm3\sigma$ besides the BEM model drag
    coefficient. The intermittency in this range in the stochastic
    model drag coefficient is mostly covered by Gaussian fluctuations.
    The strong intermittency in the BEM model drag coefficient as well
    as slight to pronounced intermittency in other force coefficients
    in both the stochastic and classical BEM model cases can probably
    be addressed to the same reason described above. The magnitude of
    the stochastic model force standard deviations is larger than the
    classical BEM model forces; compare Tables
    \ref{Table:Mod-kurtosis-sd} and \ref{Table:Static-kurtosis-sd}.
    The reason is the extended local force dynamics in case of
    stochastic model. The behavior of the standard deviations of the
    increment signals between stochastic and classical BEM models is
    very similar as observed between stochastic and dynamic stall
    models.

As described in Section \ref{subsec-Num-setup} the stochastic model
parameters have necessarily been obtained under different conditions
than the simulations carried out here. As a first compensation, the
diffusion function has been scaled according to the interpolated
hub-height turbulence intensity of wind input taken for the present
computations. This can; however, not guarantee completely comparable
flow situations. Therefore, the comparability of results between the
stochastic model and the dynamic stall and BEM models is limited at
the current state. For better quantitative comparison of stochastic
versus dynamic stall and BEM models, in the future the stochastic
model parameters have to be derived from measurements in more
realistic flow situations at wind turbine airfoils. Also a comparison
to load measurements at a real WEC should be performed.

Additionally, at the present state, the stochastic model could not
generate the expected intermittency in the resulting forces, which is
probably washed-out by the Gaussian noise term of the model. To
improve this point, multiplicative noise may be introduced to achieve
intermittency in the forces corresponding to intermittent
properties of atmospheric flows, which is currently work in progress.

The stochastic model is being developed to extract and provide more
complete local loading information on wind turbine blades, which
could lead to an optimized rotor design under turbulent wind inflow.
Here we have shown that the local force dynamics can be provided by a
stochastic model integrated into existing BEM codes used by
aerodynamic models such as AeroDyn. Further work will have to include
more realistic model parameterization and corresponding experiments to
allow for quantitative evaluation of the results.

\begin{acknowledgments}
  We appreciate the open access to the FAST and AeroDyn archives
  granted by the NREL team. M.~R.~Luhur kindly acknowledges financial
  support for higher studies by Quaid-e-Awam University of
  Engineering, Sciences and Technology, Nawabshah, Pakistan.
\end{acknowledgments}

\bibliographystyle{asmems4}
\bibliography{asme2e}

\begin{thebibliography}{10}

\bibitem{GWEC13}
GWEC, 2013.
\newblock {Global Wind Report-Annual Market Update 2012}.
\newblock Technical report, Global Wind Energy Council, Brussels, Belgium,
  April.

\bibitem{Vermeer03}
Vermeer, L.~J., S{\o}rensen, J.~N., and Crespo, A., 2003.
\newblock ``Wind turbine wake aerodynamics''.
\newblock {\em Progress in Aerospace Sciences, {\bf 39}}(6--7),
  August--October, pp.~467--510. doi: 10.1016/S0376--0421(03)00078--2.

\bibitem{Liu12}
Liu, S., and Janajreh, I., 2012.
\newblock ``Development and application of an improved blade element momentum
  method model on horizontal axis wind turbines''.
\newblock {\em International Journal of Energy and Environmental Engineering,
  {\bf 3}}, October, pp.~30 (10 pages). doi: 10.1186/2251--6832--3--30.

\bibitem{Ahlund04}
{\AA}hlund, K., 2004.
\newblock {Investigation of the NREL NASA/Ames Wind Turbine Aerodynamics
  Database}.
\newblock Scientific report FOI-R--1243--SE, Swedish Defence Research Agency,
  Stockholm, Sweden, June.

\bibitem{Hansen11}
Hansen, M. O.~L., and Madsen, H.~A., 2011.
\newblock ``Review paper on wind turbine aerodynamics''.
\newblock {\em Journal of Fluids Engineering, {\bf 133}}(11), October,
  p.~114001 (12 pages). doi: 10.1115/1.4005031.

\bibitem{Jonkman05}
Jonkman, J.~M., and { Buhl Jr.}, M.~L., 2005.
\newblock {FAST User's Guide}.
\newblock Technical report NREL/EL-500-38230, National Renewable Energy
  Laboratory, Golden, Colorado, USA, August.

\bibitem{Laino03}
Laino, D.~J., and Hansen, A.~C., 2003.
\newblock {User's Guide to the Wind Turbine Dynamics Computer Program YawDyn}.
\newblock Technical report prepared for NREL under subcontract No.
  TCX-9-29209-01, University of Utah, Salt Lake City, USA, January.

\bibitem{Laino01}
Laino, D.~J., and Hansen, A.~C., 2001.
\newblock {User's Guide to the Computer Software Routines AeroDyn Interface for
  ADAMS}.
\newblock Technical report prepared for NREL under subcontract No.
  TCX-9-29209-01, University of Utah, Salt Lake City, USA, September.

\bibitem{Mulski12}
Mulski, S., 2012.
\newblock ``Simpack multi-body simulation''.
\newblock In Proceedings of the Wind and Drivetrain Conference, Hamburg,
  Germany, SIMPACK AG.

\bibitem{Buhl06}
{Buhl Jr.}, M.~L., and Manjock, A., 2006.
\newblock {A Comparison of Wind Turbine Aeroelastic Codes Used for
  Certification}.
\newblock Conference paper NREL/CP-500-39113, National Renewable Energy
  Laboratory, Golden, Colorado, USA, January.

\bibitem{Oye99}
{\O}ye, S., 1999.
\newblock {FLEX5 User Manual}.
\newblock Technical report, Danske Techniske Hogskole.

\bibitem{Moriarty05}
Moriarty, P.~J., and Hansen, A.~C., 2005.
\newblock {AeroDyn Theory Manual}.
\newblock Technical report NREL/EL-500-36881, National Renewable Energy
  Laboratory, Golden, Colorado, USA, January.

\bibitem{Weinzierl11}
Weinzierl, G., 2011.
\newblock ``{A BEM Based Simulation-Tool for Wind Turbine Blades with Active
  Flow Control Elements.}''.
\newblock {Diploma Thesis}, Technical University of Berlin, Berlin, Germany,
  April.

\bibitem{Luhur14}
Luhur, M.~R., Peinke, J., Schneemann, J., and W{\"a}chter, M., 2014.
\newblock ``Stochastic modeling of lift and drag dynamics under turbulent wind
  inflow conditions''.
\newblock {\em Wind Energy}, doi: 10.1002/we.1699, published online.

\bibitem{Laino02}
Laino, D.~J., and Hansen, A.~C., 2002.
\newblock {User's Guide to the Wind Turbine Aerodynamics Computer Software
  AeroDyn}.
\newblock Technical report prepared for NREL under subcontract No.
  TCX-9-29209-01, University of Utah, Salt Lake City, USA, December.

\bibitem{Kelley07}
Kelley, N.~D., and Jonkman, B.~J., 2007.
\newblock {Overview of the TurbSim Stochastic Inflow Turbulence Simulator}.
\newblock Technical report NREL/TP-500-41137, National Renewable Energy
  Laboratory, Golden, Colorado, USA, April.

\bibitem{Jonkman12}
Jonkman, B.~J., and Kilcher, L., 2012.
\newblock {TurbSim User's Guide: Version 1.06.00}.
\newblock Technical report draft version, National Renewable Energy Laboratory,
  Golden, Colorado, USA, September.

\bibitem{Mendez06}
M{\'e}ndez, J., and Greiner, D., 2006.
\newblock ``Wind blade chord and twist angle optimization by using genetic
  algorithms''.
\newblock In Proceedings of the Fifth International Conference on Engineering
  Computational Technology, B.~Topping, G.~Montero, and R.~Montenegro, eds.,
  Las Palmas de Gran Canaria, Spain, Civil-Comp Press.

\bibitem{Note1}
Note1.
\newblock The azimuth angle describes the blade angular position in one cycle
  measured in clockwise direction such that it is zero when the blade is
  pointing vertically downwards.

\bibitem{Manwell09}
Manwell, J.~F., McGowan, J.~G., and Rogers, A.~L., 2009.
\newblock {\em Wind Energy Explained: Theory, Design and Application},
  $2^{nd}$~ed.
\newblock John Wiley \& Sons, Chichester, UK.

\bibitem{Glauert35}
Glauert, H., 1935.
\newblock {\em Airplane Propellers}, aerodynamic theory w.f. durand~ed.
\newblock Springer, Berlin, Germany.

\bibitem{Buhl05}
{Buhl Jr.}, M.~L., 2005.
\newblock {A new empirical relationship between thrust coefficient and
  induction factor for the turbulent windmill state}.
\newblock Technical report NREL/TP-500-36834, National Renewable Energy
  Laboratory, Golden, Colorado, USA, August.

\bibitem{Glauert26A}
Glauert, H., 1926.
\newblock {\em {A General Theory of the Autogyro}}, volume 1111 of reports and
  memoranda~ed.
\newblock British ARC, UK.

\bibitem{Glauert26B}
Glauert, H., 1926.
\newblock {\em { The Analysis of Experimental Results in the Windmill Brake and
  Vortex Ring States of an Airscrew}}, volume 1026 of reports and memoranda~ed.
\newblock HMSO, London, UK.

\bibitem{Pitt81}
Pitt, D.~M., and Peters, D.~A., 1981.
\newblock ``Theoretical prediction of dynamic-inflow derivatives''.
\newblock {\em Vertica, {\bf 5}}(1), March, pp.~21--34.

\bibitem{Coleman45}
Coleman, R.~P., Feingold, A.~M., and Stempin, C.~W., 1945.
\newblock {Evaluation of the Induced-Velocity Field of an Idealized Helicopter
  Rotor}.
\newblock Wartime report NACA ARR L5E10, National Advisory Committee for
  Aeronautics, Washington, USA, June.

\bibitem{Burton11}
Burton, T., Sharpe, D., Jenkins, N., and Bossanyi, E., 2011.
\newblock {\em Wind Energy Handbook}, $2^{nd}$~ed.
\newblock John Wiley \& Sons, Chichester, UK.

\bibitem{Risken96}
Risken, H., 1996.
\newblock {\em The Fokker-Planck Equation}, $2^{nd}$~ed.
\newblock Springer, Berlin, Germany.

\bibitem{Cordle10}
Cordle, A., 2010.
\newblock {State-of-the-art in design tools for floating offshore wind
  turbines}.
\newblock Deliverable report under contract No. 019945 (SES6), UpWind project,
  Bristol, UK, March.

\bibitem{Stoevesandt09}
Stoevesandt, B., and Peinke, J., 2009.
\newblock ``Changes in angle of attack on blades in the turbulent wind field''.
\newblock In Proceedings of EWEC 2009, Parc Chanot, Marseille, France.

\bibitem{Note2}
Note2.
\newblock The appearance of harmonics of the 1P period seems to be typical for
  the rotating frame of reference of the rotor \cite {Jonkman2014}.

\bibitem{Morales12}
Morales, A., W{\"a}chter, M., and Peinke, J., 2012.
\newblock ``Characterization of wind turbulence by higher order statistics''.
\newblock {\em Wind Energy, {\bf 15}}(3), April, pp.~391--406. doi:
  10.1002/we.478.

\bibitem{Jonkman2014}
Jonkman, J., 2014.
\newblock Personal communication, National Renewable Energy Laboratory, USA.

\end{thebibliography}
\end{document}